\documentclass{article}
\pdfoutput=1
\usepackage{amsmath,amssymb,gensymb,color,graphicx,array,authblk,multirow}
\usepackage{natbib,hyperref,aas_macros}
\usepackage[margin=22mm]{geometry}
\usepackage{listings}
\usepackage[utf8]{inputenc}
\newcommand{\code}[1]{\texttt{#1}}
\renewcommand{\Re}{\operatorname{Re}}
\renewcommand{\Im}{\operatorname{Im}}
\newcommand{\rtwo}{\frac{1}{\sqrt{2}}}
\newcommand{\rtd}{\frac{\delta^+}{\sqrt{2}}}
\newcommand{\rtc}{\frac{\delta^-}{\sqrt{2}}}
\newcommand{\I}{\tilde I}
\newcommand{\Q}{\tilde Q}
\newcommand{\U}{\tilde U}
\newcommand{\V}{\tilde V}
\newcommand{\J}{{\tilde E}}
\renewcommand{\vec}[1]{{\bf #1}}

\DeclareFixedFont{\ttb}{T1}{txtt}{bx}{n}{9} % for bold
\DeclareFixedFont{\ttm}{T1}{txtt}{m}{n}{9}  % for normal
\definecolor{deepblue}{rgb}{0,0,0.5}
\definecolor{deepred}{rgb}{0.6,0,0}
\definecolor{deepgreen}{rgb}{0,0.5,0}
\definecolor{border}{rgb}{0.9,0.9,0.9}
\title{Time-ordered data simulation and map-making for the PIXIE Fourier transform spectrometer}

\author[1,2]{S. K. Næss}
\author[2,3,4]{J. Dunkley}
\author[5]{A. Kogut}
\author[5,6]{D. J. Fixsen}
\affil[1]{Center for Computational Astrophysics, Flatiron Institute}
\affil[2]{Oxford Astrophysics, Keble Road, Oxford, OX1 3RH, UK}
\affil[3]{Princeton Physics, Jadwin Hall, Washington Road, Princeton NJ 08544}
\affil[4]{Princeton Astrophysics, Peyton Hall, Ivy Lane, Princeton NJ 08544}
\affil[5]{Goddard Space Flight Center}
\affil[6]{University of Maryland}

\begin{document}
\maketitle
\begin{abstract}
	We develop a time-ordered data simulator and map-maker for the
	proposed PIXIE Fourier transform spectrometer and use them to
	investigate the impact of polarization leakage, imperfect collimation,
	elliptical beams, sub-pixel effects, correlated noise and
	spectrometer mirror jitter on the PIXIE data analysis. We find
	that PIXIE is robust to all of these effects, with the exception
	of mirror jitter which could become the dominant source of noise
	in the experiment if the jitter is not kept significantly below
	$0.1\rm\mu m\sqrt{s}$. Source code is available at \url{https://github.com/amaurea/pixie}.
\end{abstract}

\section{Introduction}
In 1989 the COBE experiment fielded two instruments that would revolutionize the
study of the cosmic microwave background (CMB): the differential microwave
radiometer (DMR), which provided the first measurement of the angular power
spectrum \citep{cobe-dmr-1992}; and the far-infrared absolute spectrophotometer (FIRAS), which
measured its frequency spectrum and showed it to be blackbody to exquisite
precision \citep{cobe-firas-1996}. Since then DMR has been succeeded by a large number of experiments
that have improved the angular power spectrum by several orders of magnitude
both in sensitivity and angular resolution \citep{bennett/etal/2013, planck_mission/2013, bicep2-planck, act-2017, spt-2017}. However, the nearly 30 year old
FIRAS result remains our best measurement of the CMB frequency spectrum.

The Primordial Inflation Explorer (PIXIE) is a proposed successor to FIRAS,
with $\sim 1000$ times higher sensitivity, polarization support, 4 times
higher angular resolution and reduced systematics.
It would provide $1.6\degree$ FWHM full-sky maps in Stokes I, Q and U parameters
in 480 frequency channels from 15 GHz to 7 THz (though it would be noise dominated
beyond ca. 4 THz) to a depth of $70\rm nK\degree$ (that is, a square degree
variance of $1400~nK^2$) after 4 years of integration
\citep{pixie2011}. This corresponds to providing a frequency spectrum 10 times more
sensitive than FIRAS CMB monopole spectrum in each $1\degree$ pixel of the sky.
That is sufficient to constrain the spectral distortion parameters to $\mu < 4\cdot 10^{-7}$
(probing the ultra-small-scale primordial power spectrum and exotic pre-recombination
particle decay) \citep{pixie-forecast} and $y < 2\cdot 10^{-9}$; to measure the optical depth at reionization
to $\sigma(\tau) = 0.002$ (essential for getting a robust neutrino mass measurement);
and to measure the tensor-to-scalar ratio to $\sigma(r) < 0.001$ \citep{pixie-s4}.
And crucially, PIXIE's dense and broad frequency coverage would allow for robust
foreground separation, especially dust.

To make use of these huge increases in sensitivity, a corresponding reduction in
systematic errors is needed. PIXIE's systematics were studied in detail by
\citet{pixie-systematics}, who concluded that any residual errors after corrections
would be at the sub nK level, far below the instrumental noise. However, so far
no end-to-end simulations of PIXIE have been performed.

In this paper we present a python-based time-ordered data simulator and
map-maker for PIXIE, based on the mission concept proposed to NASA in 2016, and
use them to make PIXIE spectral sky maps. We then use this framework to study
the impact of some of the relevant systematic effects.

\section{The PIXIE Fourier transform spectrometer}
\begin{figure}
	\centering
	\includegraphics[angle=90,width=170mm]{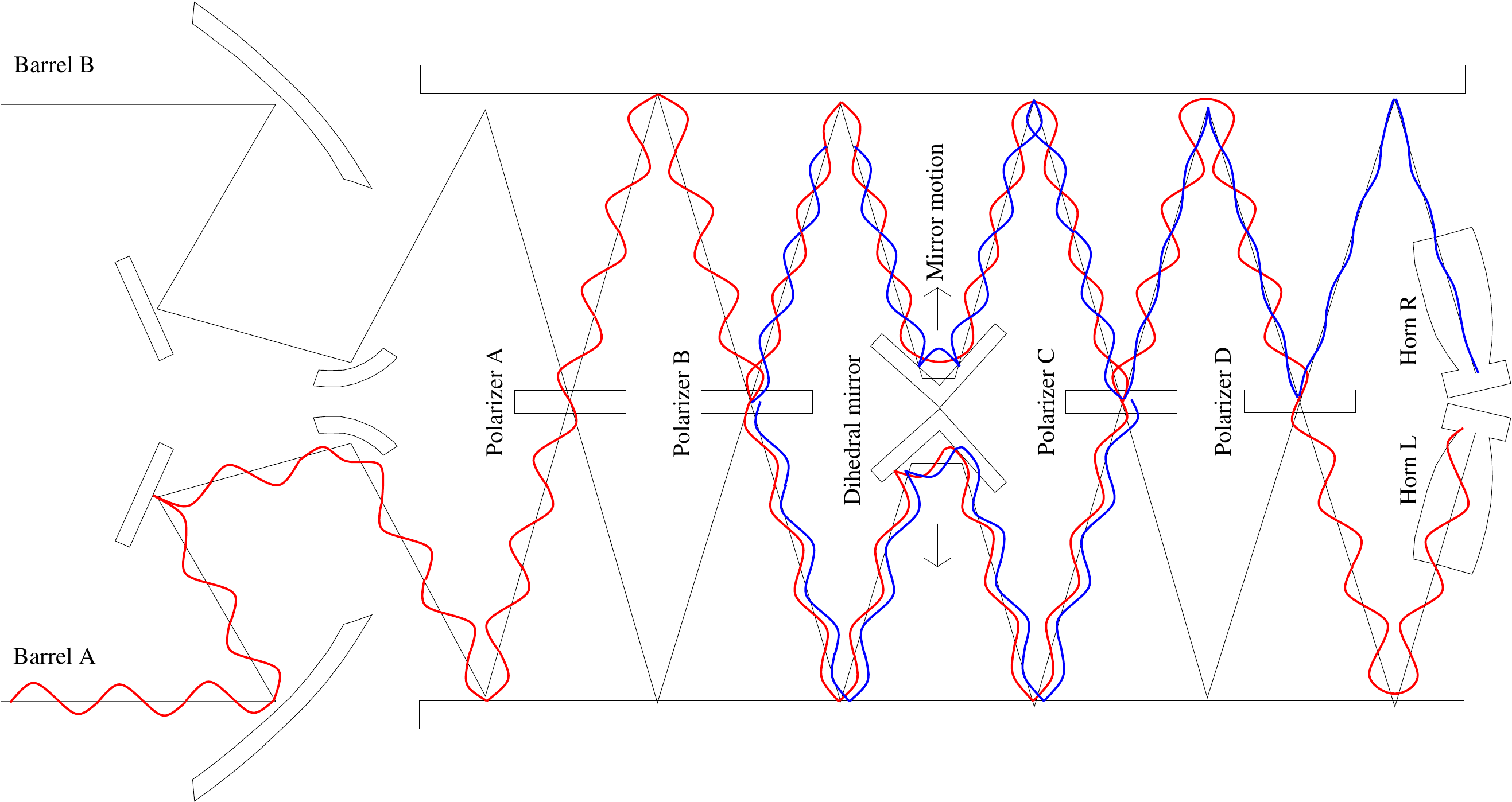}
	\caption{PIXIE optics example. Monochromatic pure vertical polarization (red) enters barrel
	A at the left side of the figure. It passes through the vertical polarizer A
	unmolested, but is split into two different linear combinations of vertical
	and horizontal (blue) polarization at the diagonal polarizer B. It then encounters
	the moving mirror, which in this example slightly retards the top optical path.
	The rays are combined in the next diagonal polarizer C, but due to the
	phase delay the horizontal polarization does not fully cancel. The vertical
	polarization is correspondingly lower. The final vertical polarizer (D)
	sends the vertical/horizontal polarization to horn L/R respectively.
	If the mirror displacement were zero, all radiation entering barrel A/B
	would end up in horn L/R.}
	\label{fig:fts}
\end{figure}
Unlike most CMB experiments, but like FIRAS, PIXIE does not use a
large number of detectors and frequency filters to measure the
frequency spectrum. Instead it splits the incoming radiation into
two paths, introduces a variable delay between them, and then lets them
interfere with each other. For each delay used, this results in
a measurement of the radiation's autocorrelation function, which
once mapped out can be Fourier transformed to recover the frequency
spectrum. This indirect way of measuring the spectrum is called
a Fourier transform spectrometer (FTS).

To improve the dynamic
range PIXIE uses two FTSes (labeled ``A'' and ``B'' in the following)
each with its own opening (``barrel'') towards either the sky or a
reference blackbody. The two barrels are collimated, and the
interferometers are coupled in a total-intensity-nulling configuration.
The situation is illustrated in figure~\ref{fig:fts}.

The barrels are operated in one of two modes:
In \emph{single-barrel mode} one barrel is
exposed to the sky while the other is covered by a 2.725 K calibrator
\citep{pixie2011}.
This cancels the contribution from the CMB monopole, leaving the
3 orders of magnitude smaller dipole as the dominant signal.
In \emph{double-barrel mode} both barrels are exposed to the sky.
In this case the entire total intensity signal is canceled, and
only the much weaker polarization signal is left.
In either case, the nulling greatly reduces the amplitude of the
modulated signal that reaches the detectors. Many systematic effects
are proportional to the total signal, and are therefore similarly
reduced by this technique.

To be able to simulate PIXIE observations, we need to know how
the power incident on its detectors relates to the incoming radiation.
This was done by \cite{pixie2011}, but is repeated here in more detail
for convenience.

We can expand the electric field $\vec E^A(t)$,
$\vec E^B(t)$ that enters PIXIE's barrels in terms of Jones vectors as
\begin{align}
	\vec E^A(t) &= \Re \int_0^\infty d\omega \Big(\J^A_x(\omega)\vec e_x +
		\J^A_y(\omega)\vec e_y\Big) e^{i(kz-\omega t)} \notag \\
	\vec E^B(t) &= \Re \int_0^\infty d\omega \Big(\J^B_x(\omega)\vec e_x +
		\J^B_y(\omega)\vec e_y\Big) e^{i(kz-\omega t)}
\end{align}
where $\omega$ is the angular frequency of the radiaton and $\vec \J^A(\omega)$
and $\vec \J^B(\omega)$ are (complex) Jones vectors at that
angular frequency.

After entering the barrels the light encounters polarizer A, which lets through
vertical polarization and reflects horizontal\footnote{
	Before this it encounters the primary mirror, folding flats, secondary mirror,
	and transfer mirror 1, but these lead to the same phase shifts on both
	the A and B side optical paths, so they can be neglected.}. After this,
the Jones vectors in left (A) and right (B) shafts are
\begin{align}
	\vec \J^{A1} &= \J^A_x \vec e_x + \J^B_y \vec e_y &
	\vec \J^{B1} &= \J^B_x \vec e_x + \J^A_y \vec e_y
\end{align}
After passing through the diagonal polarizer B, we have
\begin{align}
	\vec \J^{A2} &= \rtwo [\J^A_x+\J^B_y]\vec e_a + \rtwo[-\J^B_x+\J^A_y]\vec e_b &
	\vec \J^{B2} &= \rtwo [\J^B_x+\J^A_y]\vec e_a + \rtwo[-\J^A_x+\J^B_y]\vec e_b
\end{align}
where $\vec e_a \equiv \rtwo [\vec e_x + \vec e_y]$ and $\vec e_b
\equiv \rtwo [-\vec e_x + \vec e_y]$. The dihedral mirror then
imparts a path length difference between the two sides, advancing
A by $\frac12\Delta t$ and retarding B by $\frac12\Delta t$, which
is achieved by multiplying A by $\delta^+ = e^{-\frac12 i\omega\Delta t}$
and B by $\delta^- = e^{\frac12 i\omega\Delta t}$:
\begin{align}
	\vec \J^{A3} &= \rtd [\J^A_x+\J^B_y]\vec e_a + \rtd[-\J^B_x+\J^A_y]\vec e_b &
	\vec \J^{B3} &= \rtc [\J^B_x+\J^A_y]\vec e_a + \rtc[-\J^A_x+\J^B_y]\vec e_b
\end{align}
Polarizer C is also diagonal, resulting in
\begin{align}
	\vec \J^{A4} &= \rtd [\J^A_x+\J^B_y]\vec e_a + \rtc [-\J^A_x+\J^B_y]\vec e_b &
	\vec \J^{B4} &= \rtc [\J^B_x+\J^A_y]\vec e_a + \rtd [-\J^B_x+\J^A_y]\vec e_b
\end{align}
And the final polarizer D is vertical. The output of this enters
the left (L) and right (R) feedhorns.
\begin{align}
	\vec \J^L = \vec \J^{A5} &= \frac12 \Big[\delta^+(\J^A_x+\J^B_y)+\delta^-(\J^A_x-\J^B_y)\Big]\vec e_x
		+ \frac12 \Big[\delta^-(\J^B_x+\J^A_y)+\delta^+(-\J^B_x+\J^A_y)\Big]\vec e_y \notag \\
		&= \big[\J^A_x\cos(\omega\Delta t/2) - i\J^B_y\sin(\omega\Delta t/2)\big]\vec e_x
		+  \big[\J^A_y\cos(\omega\Delta t/2) + i\J^B_x\sin(\omega\Delta t/2)\big]\vec e_y \\
	\vec \J^R = \vec \J^{B5} &= \frac12 \Big[\delta^-(\J^B_x+\J^A_y)+\delta^+(\J^B_x-\J^A_y)\Big]\vec e_x
		+ \frac12 \Big[\delta^+(\J^A_x+\J^B_y)+\delta^-(-\J^A_x+\J^B_y)\Big]\vec e_y \notag \\
		&= \big[\J^B_x\cos(\omega\Delta t/2) + i\J^A_y\sin(\omega\Delta t/2)\big]\vec e_x
		+  \big[\J^B_y\cos(\omega\Delta t/2) - i\J^A_x\sin(\omega\Delta t/2)\big]\vec e_y
\end{align}

\subsection{Stokes parameters}
\label{sect:stokes}
After passing through all this, the light enters the feedhorns and
hits the detectors. The power deposited here can be decomposed into
Stokes parameters\footnote{The quantities with tildes are for a single plane wave.
The full Stokes parameters are obtained by integrating these. E.g. $I(\Delta t) =
	\int_0^\infty \I(\omega) d\omega$.}
\begin{align}
	\I &= \langle |\J_x|^2\rangle + \langle|\J_y|^2\rangle &
	\Q &= \langle |\J_x|^2\rangle - \langle|\J_y|^2\rangle &
	\U &= 2\Re\langle \J_x\J_y^*\rangle &
	\V &= -2\Im\langle \J_x\J_y^*\rangle
\end{align}
so we need to evaluate $P_{xx} = \langle |\J_x|^2\rangle$,
$P_{yy} = \langle |\J_y|^2\rangle$ and $P_{xy} = \langle \J_x\J_y^*\rangle$.
For the left horn\footnote{The right horn follows by symmetry: $(L,A,B)\leftrightarrow
(R,B,A)$.} we get
\begin{align}
	P^L_{xx} &= \langle \J^L_x\J^{L*}_x \rangle \notag \\
		&= \frac12\big[1+\cos(\omega\Delta t)\big]\langle\J^A_x\J^{A*}_x\rangle
		+ \frac12\big[1-\cos(\omega\Delta t)\big]\langle\J^B_y\J^{B*}_y\rangle
		- \frac{i}2\langle\J^A_x\J^{B*}_y + \J^{A*}_x\J^B_y\rangle\sin(\omega\Delta t) \notag \\
		&= \frac14\big[\I^A+\I^B+\Q^A-\Q^B + (\I^A-\I^B+\Q^A+\Q^B)\cos(\omega\Delta t) -
		4\Im(\J^A_x\J^{B*}_y)\sin(\omega\Delta t) \big] \\
	P^L_{yy} &= \frac14\big[\I^B+\I^A+\Q^B-\Q^A - (\I^B-\I^A+\Q^B+\Q^A)\cos(\omega\Delta t) -
		4\Im\langle\J^B_x\J^{A*}_y\rangle\sin(\omega\Delta t) \big ]\\
	P^L_{xy} &= \frac14\big[\U^A-\U^B - i\V^A-i\V^B + (\U^A+\U^B -i\V^A+i\V^B)\cos(\omega\Delta t)
		+2i\langle\J^A_x\J^{B*}_x+\J^B_y\J^{A*}_y\rangle\sin(\omega\Delta t)
\end{align}
Hence
\begin{align}
	\I^L &= \frac12\big[\I^A+\I^B+(\I^A-\I^B)\cos(\omega\Delta t)
		-2\Im\langle\J^A_x\J^{B*}_y+\J^B_x\J^{A*}_y\rangle\sin(\omega\Delta t) \big] \notag \\
	\Q^L &= \frac12\big[\Q^A-\Q^B+(\Q^A+\Q^B)\cos(\omega\Delta t)
		-2\Im\langle\J^A_x\J^{B*}_y-\J^B_x\J^{A*}_y\rangle\sin(\omega\Delta t) \big] \notag \\
	\U^L &= \frac12\big[\U^A-\U^B+(\U^A+\U^B)\cos(\omega\Delta t)
		-2\Im\langle\J^A_x\J^{B*}_x+\J^B_y\J^{A*}_y\rangle\sin(\omega\Delta t) \big] \notag \\
	\V^L &= \frac12\big[\V^A+\V^B+(\V^A-\V^B)\cos(\omega\Delta t)
		-2\Re\langle\J^A_x\J^{B*}_x+\J^B_y\J^{A*}_y\rangle\sin(\omega\Delta t) \big]
\end{align}
The value of the barrel cross-terms depends on whether PIXIE is in single
or double barrel mode.

\paragraph{Single barrel mode}
In single barrel mode only one barrel is exposed to the sky; the other one
observes a static calibrator object. The light entering the two barrels is
therefore uncorrelated, and all the cross-terms disappear.
\begin{align}
	\I^L &= \frac12\big[\I^A+\I^B+(\I^A-\I^B)\cos(\omega\Delta t)\big] &
	\Q^L &= \frac12\big[\Q^A-\Q^B+(\Q^A+\Q^B)\cos(\omega\Delta t)\big] \notag \\
	\V^L &= \frac12\big[\V^A+\V^B+(\V^A-\V^B)\cos(\omega\Delta t)\big] &
	\U^L &= \frac12\big[\U^A-\U^B+(\U^A+\U^B)\cos(\omega\Delta t)\big] \label{eq:single}
\end{align}

\paragraph{Double barrel mode}
In double barrel mode the two barrels are both coaligned and exposed to the sky,
so they observe the same wavefront entering. As PIXIE's angular resolution is
not infinite it is sensitive to wavefronts that are off-axis by a few degrees,
causing the two barrels to act as a 2-element spatial interferometer.
Light arriving from a single direction $\hat n$ will hit Barrel B a time
$\tau = \hat n \cdot \vec b / c$ before barrel A, where $\vec b$ is the distance vector from
barrel A to barrel B, and $c$ is the speed of light (see~\citet[appendix]{pixie-systematics}).
So in this case $\vec \J^B = \gamma \vec\J^A$ with $\gamma = e^{-i\omega\tau}$.
\begin{align}
	\I^L &= \I^A+\V^A\cos(\omega\tau)\sin(\omega\Delta t) &
	\Q^L &= \Q^A\cos(\omega\Delta t)-\U^A\sin(\omega\tau)\sin(\omega\Delta t) \notag \\
	\V^L &= \V^A-I^A\cos(\omega\tau)\sin(\omega\Delta t) &
	\U^L &= \U^A\cos(\omega\Delta t)-\Q^A\sin(\omega\tau)\sin(\omega\Delta t) \label{eq:double}
\end{align}
The total signal will be the contribution from all directions integrated over the
barrel beam. The cross terms (those proportional to $\sin(\omega\Delta t)$ are
antisymmetric with respect to the baseline $\vec b$ separating the two barrels, making it cancel
to first order when integrated over the symmetric beam. Furthermore, the
$\sin(\omega \Delta t)$ dependence of any residual is anti-symmetric
with respect to the mirror stroke, forcing these terms into the (unphysical) imaginary part of
the frequency maps. This imaginary part contains none of the real signal and would usually
be discarded, but can be inspected as a test for systematic errors.

If the barrels are not perfectly collimated, or if they have asymmetric
sidelobes or different beam size, then the situation will be more complicated,
as only part of the radiation that enters the barrels will be correlated.

These cross terms are not implemented in the current version of the simulator,
but for the reasons above we do not expect this to impact our results meaningfully.

In the absence of the cross terms terms, the signal in double barrel mode is
identical to that of single barrel mode with $(I^A,Q^A,U^A,V^A)=(I^B,Q^B,U^B,V^B)$,
so we can use the single barrel eqs~(\ref{eq:single}) in the following
without further loss of generality.

A fuller treatment of the effects of beam asymmetries can be found in
\citet{pixie-multi-moded-beams-2018}.

\subsection{Detector response}
PIXIE has an x and y-oriented detector in each horn. The power deposited on each
of these is
\begin{align}
	s^L_x(\Delta t)
		&= \frac14 \int_0^\infty\Big(\I^A+\I^B+\Q^A-\Q^B + (\I^A-\I^B+\Q^A+\Q^B)\cos(\omega\Delta t)\Big)d\omega \notag \\
		&= \frac14 \big[I^A+I^B+Q^A-Q^B\big]_0 + \frac14\big[I^A-I^B+Q^A+Q^B\big]_{\Delta t} \notag \\
	s^L_y(\Delta t)
		&= \frac14 \big[I^B+I^A+Q^B-Q^A\big]_0 - \frac14\big[I^B-I^A+Q^B+Q^A\big]_{\Delta t} \notag \\
	s^R_x(\Delta t)
		&= \frac14 \big[I^B+I^A+Q^B-Q^A\big]_0 + \frac14\big[I^B-I^A+Q^B+Q^A\big]_{\Delta t} \notag \\
	s^R_y(\Delta t)
		&= \frac14 \big[I^A+I^B+Q^A-Q^B\big]_0 - \frac14\big[I^A-I^B+Q^A+Q^B\big]_{\Delta t}
\end{align}
where all the quantities depend on $\Delta t$ and potentially $\tau$,
and where these are total Stokes parameters, not the per-frequency ones,
e.g. $I = \int_0^\infty \I(\omega) d\omega$, and $[\ldots]_{\Delta t}$
means that the quantities within should be evaluated at the time delay in
the subscript.

Combining this with the effect of PIXIE's pointing on the sky,
we can express the total detector response as a function of the
sky autocorrelation functions.
\begin{align}
\vec s^\textrm{det}(t) &= \overbrace{\frac12\begin{bmatrix}
	\vec e_I+\vec e_Q & 0 \\
	\vec e_I-\vec e_Q & 0 \\
	0 & \vec e_I+\vec e_Q  \\
	0 & \vec e_I-\vec e_Q \end{bmatrix}}^\textrm{Detector response}
	\cdot
	\overbrace{\frac12\begin{bmatrix}
	1 &  M &  1 & -M \\
	M &  1 & -M &  1
	\end{bmatrix}}^\textrm{Horn response}
	\cdot
	\overbrace{\begin{bmatrix}
	R_{1t}\cdot\vec s^\textrm{sky}_A(\hat p_{1t},0) \\
	R_{2t}\cdot\vec s^\textrm{sky}_B(\hat p_{2t},0) \\
	R_{1t}\cdot\vec s^\textrm{sky}_A(\hat p_{1t},\Delta t) \\
	R_{2t}\cdot\vec s^\textrm{sky}_B(\hat p_{2t},\Delta t)
	\end{bmatrix}}^\textrm{Barrel signal}
	\label{eq:response}
\end{align}
$\vec s^\textrm{det}(t) = [s^L_x, s^L_y, s^R_x, s^R_y]$ is the vector of
detector responses at time $t$,
$\hat p_{bt}$ is the sky pointing of barrel $b$ at time $t$,
$\vec s^\textrm{sky}_b(\hat p,\Delta t)$ is the beam-smoothed,
frequency-weighted sky autocorrelation
function Stokes vectors for the given pointing and time delay as
seen by barrel $b$
(different barrels can see different skies because one barrel may
be covered by a blackbody calibrator),
$R_{bt}$ is
a matrix that rotates the polarization basis from sky to instrument
coordinates, $M = \textrm{diag}(1,-1,-1,1)$ is a matrix that flips
the sign of linear polarization,
and $\vec e_I = (1,0,0,0)$ and $\vec e_Q = (0,1,0,0)$ are Stokes
I and Q basis vectors. PIXIE's interferometry shows up in two ways here:
The sky autocorrelation function, rather than just its intensity,
is what is measured; and the barrel signal differencing in the horn
response.

\subsection{Readout}
Of course, a real instrument does not read out data with infinite
time resolution, but as a set of discrete samples, each of which is
noisy. The PIXIE hardware will also apply a bandpass filter to avoid
aliasing and suppress low-frequency noise. Taking this into account,
we model the time-ordered data as
\begin{align}
	\vec d_i = B_{ij} \int_{t_j-\Delta t/2}^{t_j+\Delta t/2} \vec s^\textrm{det}(t) \textrm{d}t + \vec n_i
\end{align}
where $i$ is the sample index, $B$ is the bandpass filter, $\Delta t$ is the sample
interval and $\vec n_i$ is the noise in sample $i$. We implement this
sample integral by using Gaussian quadrature with $N_\textrm{sub}$ sub-samples,
with a typical value of $N_\textrm{sub}$ being 9.
%See section~\ref{s:accuracy} for a discussion of the effect of the number of sub-samples.

For the bandpass filter we used a Butterworth bandpass filter,
\begin{align}
	B(f) &= \Big(1+\Big[\frac{f}{0.01\textrm{Hz}}\Big]^{-5}\Big)^{-1}
		\Big(1+\Big[\frac{f}{100\textrm{Hz}}\Big]^{5}\Big)^{-1}.
\end{align}

\subsection{Frequency response}
\label{sect:response}
PIXIE's frequency response is limited above 1-2 THz by the
roughness of the mirrors. Scattering from the mirrors
provides a gradual decrease in the coupling of the detectors
to the sky. The resulting apodized frequency response
smoothly band-limits the signal to avoid aliasing from frequencies
about the instrument's Nyquist frequency of 7 THz. Dispersion within
the FTS from the finite spread of angles within the beam
washes out the fringe amplitude at comparable frequencies.
Synthesized channels at frequencies above $\sim$6 THz thus contain
noise but no signal, providing a convenient check of the instrument
noise.

We here model the total frequency response as $\rho(\nu) =
e^{-\left[\frac{\nu}{1.5\textrm{THz}}\right]^2}$.

\section{Pointing}
\label{sec:pointing}
PIXIE would orbit at the Sun-Earth L2 point, placing it in the ecliptic, with
a heliocentric ecliptic latitude $b=0$ and longitude $l=l_0 +
360\degree\frac{t-t_0}{T_\textrm{orbit}}$\footnote{We're ignoring the
orbital eccentricity here for simplicity, but nothing in the simulation
or mapmaking relies on the longitude changing at a constant rate,
so it would be trivial to add support for eccentricity.}
with $T_\textrm{orbit} = 1\textrm{ year}$.
In addition to this orbital motion, PIXIE also scans great circles\footnote{
Getting sufficient sun shielding might require the opening angle of the scan
to be smaller than $180\degree$. This would make the scans small circles instead.
This would result in 1. the ecliptic poles no longer being covered, and 2. the
\texttt{add\_to\_sky} operation in section~\ref{sect:mapmaker} would be
somewhat more complicated. Aside from that, nothing changes.}
perpendicular to the direction towards the sun, with a linearly increasing scan angle
$\alpha_\textrm{scan} = \alpha_{\textrm{scan},0} + 360\degree \frac{t-t_0}{T_\textrm{scan}}$.
To form an actual great circle the scan axis does not
move continuously with $b$, but updates in steps after each circle has been completed:
$\alpha_\textrm{orbit} = l_0 + 360\degree\left\lfloor\frac{t-t_0}{T_\textrm{scan}}\right\rfloor\frac{T_\textrm{scan}}{T_\textrm{orbit}}$.
On top of this scanning motion, the telescope also spins rapidly around the
barrel boresight in order to modulate
the observed polarization and reject systematics: $\alpha_\textrm{spin} =
\alpha_{\textrm{spin},0} + 360\degree\frac{t-t_0}{T_\textrm{spin}}$. And finally,
while it is spinning the dihedral mirror sweeps backwards and forwards at constant
speed, varying the path length time difference in the Fourier transform spectrometer
by $\Delta t = A_\textrm{delay}\textrm{triangle}\Big(\frac{t-t_0}{T_\textrm{stroke}}\Big)$,
with $A_\textrm{delay} = 10.40303 \textrm{mm}/c$ for the purposes of this paper, but
varying somewhat by observing mode in the real experiment, and with
$\textrm{triangle}(x)$ being the triangle wave with period 1, mean 0 and a zero
crossing at $x=0$.

\begin{figure}
	\centering
	\begin{tabular}{m{80mm}m{70mm}}
		\includegraphics[width=80mm]{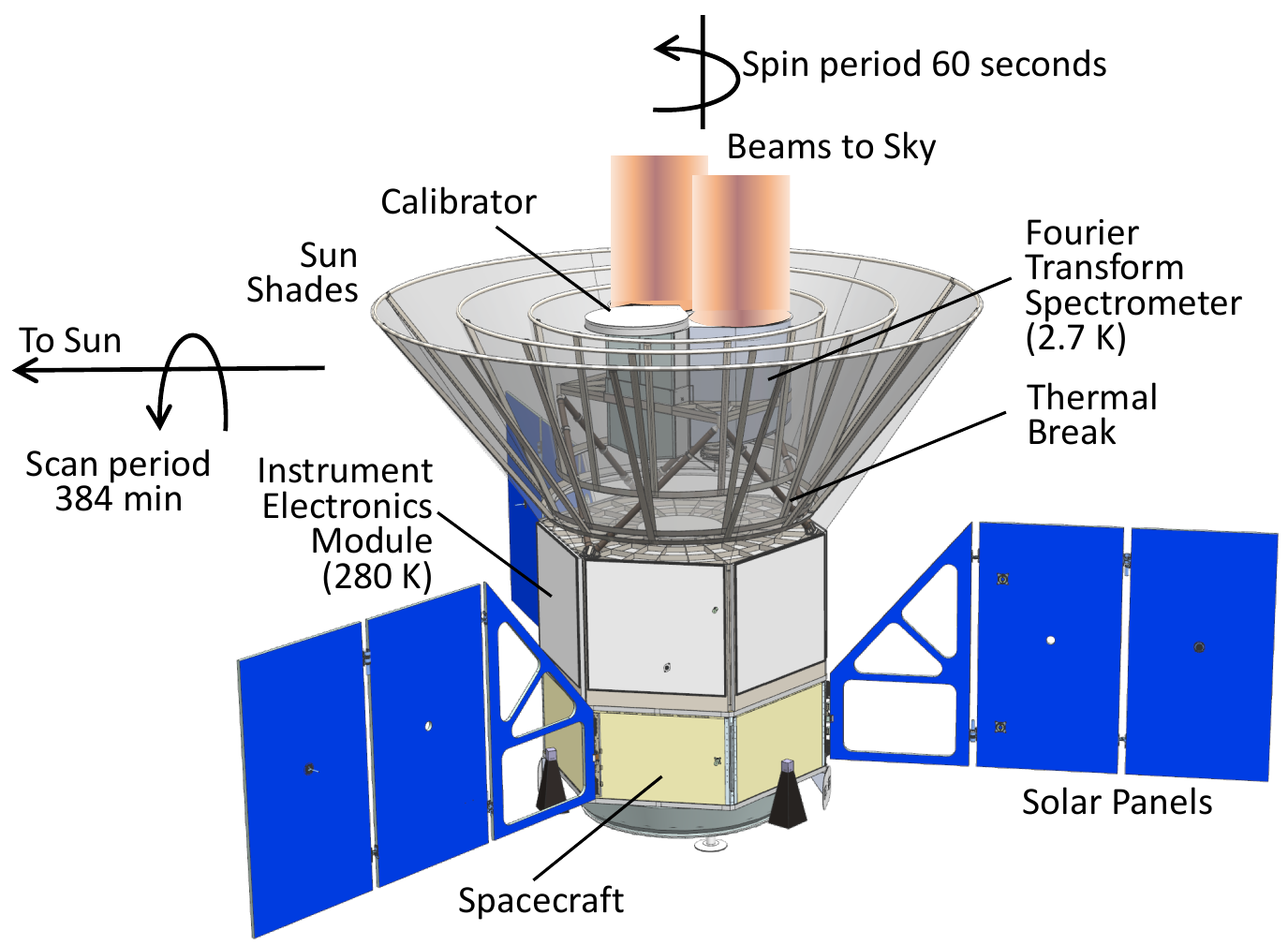} &
		\includegraphics[width=70mm,clip,trim=150mm 0 0 0]{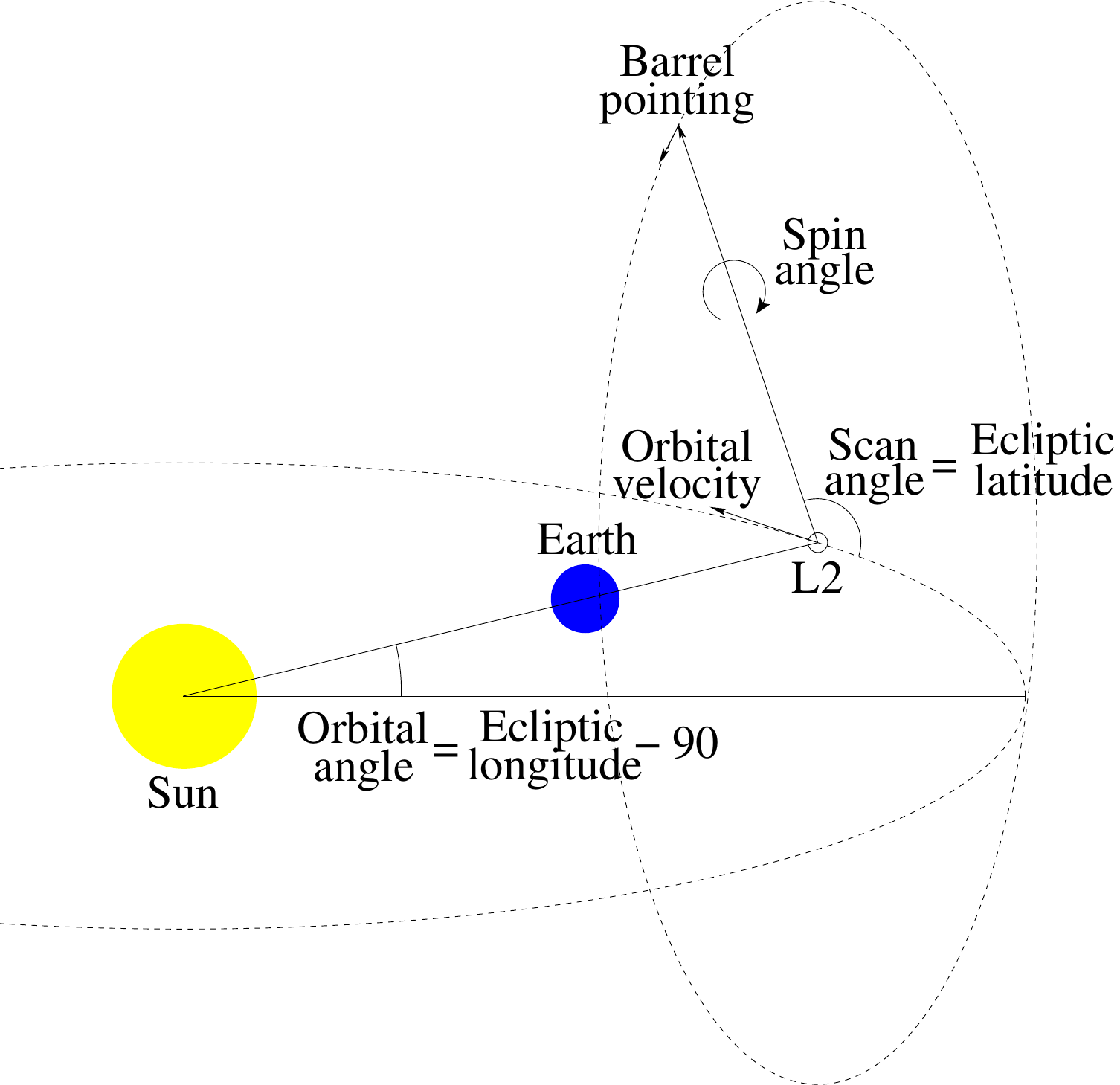}
	\end{tabular}
	\caption{\emph{Left}: The PIXIE observatory, showing the spin and scan
	axes relative to the barrel beams and the direction towards the sun.
	\emph{Right}: PIXIE would be located at the Sun-Earth L2 point,
	and would scan in great circles while pointing $90\degree$ away from the
	Sun, adjusting its orientation stepwise between each scan. This results
	in the whole sky being covered every 6 months.}
\end{figure}

To allow PIXIE mapmaking to use fast Fourier transform methods, the stroke, spin
and scan periods will be synchronized such that there is an integer number of
strokes in a spin, and an integer number of spins in a scan. We will use the
values $T_\textrm{spin} = 60 \textrm{s}$, $T_\textrm{stroke} = T_\textrm{spin}/8 =
7.5 \textrm{ sec}$, $T_\textrm{scan} = 384 T_\textrm{spin} = 384 \textrm{ min}$ here.
The time-ordered data (TOD) simulator purposefully does not depend on integer ratios to be able
to investigate the consequences of small deviations from integer ratios.

{\bf To summarize}, PIXIE moves in four different ways: It orbits with a 1 year period;
it scans in great circles perpendicular to the direction towards the sun with a 384
minute period; it spins around the boresight with a 1 minute
period; and it strokes the FTS mirror with a 7.5 second period. Aside from the orbital
period these numbers are subject to change, but the period ratios will be kept at
integer numbers.

In order to speed up our simulations we will modify the scanning pattern we
simulate in one important respect. The actual L2 orbital
period given above results in about 1370 scans per orbit, which results in
7.6 scans per PIXIE beam FWHM on the equator after half an orbit. We avoid
this oversampling by simulating a faster $T_\textrm{orbit} = 384 T_\textrm{scan}$.

A barrel-to-sky rotation matrix that implements this pointing model is
\begin{align}
	R_\textrm{tot}(b,t)  &= R_\textrm{orient}(t)R_\textrm{barrel}(b) \\
	R_\textrm{orient}(t) &= R_z\big(\alpha_\textrm{orbit}(t)\big)
		R_y\Big(\frac\pi2-\alpha_\textrm{eclip}\Big)
		R_z\big(\alpha_\textrm{scan}(t)\big)R_y\Big(\frac\pi2-\alpha_\textrm{open}\Big)
		R_z\big(\alpha_\textrm{spin}(t)\big) \\
	R_\textrm{barrel}(b) &= R_z\big(\Delta\phi(b)\big)R_y\big(\Delta\theta(b)\big)R_z\big(\Delta\psi(b)\big) \label{eq:barrel}
\end{align}

Here $R_\textrm{barrel}(b)$ represents the orientation of barrel $b$ relative
to the spacecraft. Fiducially $R_\textrm{barrel} = 1$ for both barrels, but
we include this rotation to be able to support misaligned barrels or more
complicated beams. $R_\textrm{orient}(t)$ represents PIXIE's orientation
in space at time $t$, and in addition to the angles described above includes
$\alpha_\textrm{eclip}$ and $\alpha_\textrm{open}$, which represent the
offset of PIXIE's orbital plane from the ecliptic and the opening angle
offset (to support non-great-circle scans), both of which are fiducially 0.
$R_y(\theta)$ and $R_z(\theta)$ are rotations around the $y$ and $z$ axes
by an angle $\theta$.\footnote{In ecliptic coordinates $z$ represents the
zenith and $x$ the zero longitude direction. In barrel coordinates, $z$
represents the fiducial barrel pointing.}

$R_\textrm{tot}$ encodes both the sky coordinates and polarization rotation,
\begin{align}
	x_i &= R_\textrm{tot,xi} & y_i &= R_\textrm{tot,yi} & p_i &\equiv z_i = R_\textrm{tot,zi} \\
	l        &= \tan^{-1}(p_y/p_x)  &
	b        &= \tan^{-1}\Big(\frac{p_z}{\sqrt{p_x^2+p_y^2}}\Big) &
	\gamma   &= \tan^{-1}\Big(\frac{x_z}{p_y x_x - p_x x_y}\Big).
\end{align}
with all the above being a function of the barrel index and time.
$\hat p = (p_x,p_y,p_z)$ is the pointing vector and $\gamma$ is the
polarization basis rotation, and corresponds to a Stokes rotation
matrix
\begin{align}
	R_\textrm{stok} &= \begin{bmatrix}
		1 & 0 & 0 & 0\\
		0 & \cos(2\gamma) & -\sin(2\gamma) & 0\\
		0 & \sin(2\gamma) & \cos(2\gamma) & 0 \\
		0 & 0 & 0 & 1
	\end{bmatrix}
\end{align}

%	p_{bt,i} &= R_{\textrm{tot,3i}}(b,t) \\

\section{Evaluating the sky autocorrelation function at the observed location}
As PIXIE observes the sky it mesures the autocorrelation function of the
radiation coming from the points it scans past. To simulate the PIXIE signal
we therefore need to be able to evaluate the I, Q and U autocorrelation
functions
\footnote{We're ignoring V polarization here. See section~\ref{sect:stokes}.}
at an arbitrary point $\hat p$ on the sky for an arbitrary phase
delay $\Delta t$ for each component that makes up the sky (CMB, dust, etc.).

\subsection{Precomputing the autocorrelation function as a data cube}
\label{sect:autocorr-problems}
A straightforward and general way of doing this would be to precompute the
full-sky autocorrelation function:
\begin{enumerate}
	\item Evaluate the full-sky spectrum at equi-spaced
frequencies
	\item Apply any frequency-dependent beam to each frequency map and
scale each frequency by the instrument's frequency response.
	\item Fourier transform the result to get a (pixelized version of)
		the full-sky autocorrelation function.
	\item Apply any mirror-position-dependent beam or response to
		each delay in the result.
\end{enumerate}
To read off the value at a general $(\hat p, \Delta t)$ one would then do
an interpolated lookup in this $N_\textrm{pix}$ by $N_\textrm{delay}$ data
cube. This approach has the advantage of being able to handle both frequency-
and delay-dependent beams, which are otherwise hard to implement.

However, it also has some important limitations. Because it stores the full
spectrum/autocorrelation function in each pixel, its memory requirements
scale poorly with resolution. This makes it impractical to investigate the effect
of sub-resolution features (both spectrally and spatially) - to do so would
require the data cube to be pixelized at many times higher resolution
than the PIXIE output map, which would make the memory requirements of this approach
prohibitively high. For example,
for $0.1\degree$ spatial resolution and 5000 frequency bins, storing the full-sky
autocorrelation function would need about 700 GB of RAM.

We will therefore save the data cube approach for a future investigation of the
effects of frequency- and delay-dependent beams, and use an approach that
allows for high-resolution simulations in this paper.

\subsection{Other approaches}
If one assumes a frequency-independent beam, which should be a good approximation
for PIXIE, and if the spectrum can be written as a linear sum of a smaller
number of spatial templates, then it's sufficient to apply the beam to those
templates rather than the spectrum itself. This
decouples the spatial and spectral dimensions, making it possible to evaluate the
spectrum in one pixel independently of the rest of the sky.

With this, we could
imagine the following approach: For each sample, interpolate the spectrum parameters
at $\hat p$, then evaluate the whole spectrum, apply the frequency response,
Fourier transform it,
and interpolate the value for $\Delta t$. In our example above, this would reduce
the RAM requirements by a factor of 5000. But it would introduce another prohibitive
cost: The need to evaluate the spectrum at thousands of frequencies and Fourier transform
these for every sample in the TOD.\footnote{
A hybrid approach between these two would be to precompute the autocorrelation function
for a chunk of the sky around the current sample, and reuse that for subsequent samples
until a sample falls outside the chunk, and then precompute a new chunk. We investigated
this in the hopes of being able to support frequency-dependent beams, but found that
edge effects and the flat-sky-approximation needed to perform beam-smoothing on a small
patch did not result in the required accuracy. This may still be a good approach for
frequency-independent beam simulations, though.}

\subsection{Autocorrelation by Taylor expansion}
In the end, we went for a Taylor expansion approach: The autocorrelation function
is evaluated as a perturbation around a different but similar
precomputed autocorrelation function. This is done differently for each sky component.

\subsubsection{CMB}
\label{sect:cmb}
Taking into account the instrument's frequency response $\rho(\nu)$ (see section~\ref{sect:response}),
PIXIE observes the CMB with the spectrum
\begin{align}
I^\textrm{CMB}_{\nu,I}(\hat p, \nu) &= \rho(\nu)B_\nu(\nu, T(\hat p)) = \frac{2h\nu^3\rho(\nu)}{c^2}\frac{1}{e^{\frac{h\nu}{k_BT(\hat p)}} - 1}.
\end{align}
Here the $I$ subscript indicates the Stokes intensity parameter, and $T(\hat p)$ is the
CMB temperature at pointing $\hat p$. $k_B$ and $h$ are the Boltzmann and Planck constants
respectively.
Including the Doppler dipole, T only has a contrast of order $10^{-3}$, so a Taylor
expansion in T will converge rapidly. Our goal is $<10^{-9}$ relative error, so an
expansion to 3rd order, which should give order $10^{-12}$ error, should be sufficient.
The expansion is
\begin{align}
	I^\textrm{CMB}_{\nu,I}(\hat p, \nu) &= f_0(\nu) + f_1(\nu) \Delta T + \frac12 f_2(\nu) \Delta T^2 + \frac16 f_3(\nu) \Delta T^3
\end{align}
where
\begin{align*}
	f_0 &= p(g_0-1)^{-1}    & g_0 &= e^{a/T_0} \\
	f_1 &= pf_0^2 g_1       & g_1 &= -g_0 a/T_0^2 \\
	f_2 &= p(-2 f_0 f_1 g_1 - f_0^2 g_2) &
	g_2 &= a(2 g_0-T_0 g_1)/T_0^3 \\
	f_3 &= p(-2 f_1^2 g_1 - 2f_0 f_2 g_1 - 4 f_0 f_1 g_2 - f_0^2 g_3) &
	g_3 &= -(3/T_0 + a/T_0^2)g_2 + a g_1/T_0^3
\end{align*}
and where $p = \frac{2h\nu^3\rho(\nu)}{c^2}$, $a = h\nu/k_B$ and $\Delta T(\hat p) = T(\hat p)-T_0$ with $T_0 = 2.725$K. 

The autocorrelation function is simply the cosine transform\footnote{
We implemented the cosine transform using a discrete cosine transform with a
sample interval of 0.5 GHz and a max frequency of 6.8 THz.} of the spectral power
density,
\begin{align}
	I_{\Delta t}(\Delta t) &= \int_{0}^\infty I_\nu(\nu) \cos(2\pi\nu\Delta t) d\nu
	\equiv \bar I_\nu(\Delta t).
\end{align}
where $\bar{x}$ indicates the cosine transform of $x$.
%{\color{red}This is confusing. Tilde quantities are consistently
%in frequency space in the interferometry section, but here the visually
%similar bars indiate the cosine transform of a frequency space quantity
%into a real space quantity.}
Applying this to the Taylor expansion, we get
\begin{align}
I^\textrm{CMB}_{\Delta t,I}(\hat p,\Delta t) &= \bar f_0(\Delta t) + \bar f_1(\Delta t) \Delta T + \frac12 \bar f_2(\Delta t) \Delta T^2 + \frac16 \bar f_3(\Delta t) \Delta T^3
\end{align}
Hence, we can compute the autocorrelation for any $\hat p,\Delta t$ if we simply precompute
the four position-independent functions $\{f_i\}$.\footnote{
If we had not needed to support the frequency response of the instrument,
we could have avoided the Taylor expansion by absorbing variation in $T$
into rescaling of $v$. Sadly, PIXIE has significant damping at high frequency,
so this approach does not work.} Evaluating the CMB autocorrelation at ($\hat p,\Delta t$)
is hence reduced to being able to evaluate a sampled version of $\{f_i\}$
at (non-sample) position $\Delta t$ and
the full-sky pixelized map $\Delta T$ at (non-pixel) position $\hat p$.
We perform both of these using (bi-)cubic spline interpolation
from \code{numpy.ndimage.map\_coordinates}.

The CMB has frequency-independent polarization,
so the $Q,U$ autocorrelation functions can be derived from $I$ by scaling them by
the local $Q,U$ polarization fractions. I.e. $I^\textrm{CMB}_{\Delta t,Q|U}(\hat p,\Delta t) = I^\textrm{CMB}_{\Delta t,I}(\hat p,\Delta t) \frac{I^\textrm{CMB}_{\textrm{ref},Q|U}(\hat p)}{I^\textrm{CMB}_{\textrm{ref},I}(\hat p)}$.

The input CMB map $\Delta T,Q,U$ was simulated by drawing random, Gaussian
T,E,B and $\phi$ spherical harmonics coefficients from a typical CMB power spectrum
as output by CAMB\footnote{The spectrum used is provided in the file \code{
	inputs/cl\_lensinput.dat}.} and projecting them on a sky with $0.1\degree$
pixels in equirectangular (CAR) projection using the \code{libsharp} Spherical
Harmonics Transform library \citep{libsharp}. The lensing potential $\phi$ was then used to
lens the T, Q and U maps. We then added the 2.725 K CMB monopole to the T
component before Doppler boosting the sky\footnote{$\beta=0.0012301$ towards
ecliptic coordinates $l=171.646$, $b=-11.141$.}
to account for our motion relative to the CMB, resulting in the CMB dipole.

\subsubsection{Dust}
\label{sect:dust}
We model the dust as a modified blackbody with constant
$T_D=19.6$K and $\beta=1.59$, but varying opacity. The observed spectrum is thus
\begin{align}
I^\textrm{dust}_{\nu,i}(\hat p,\nu) &= A_i(\hat p) \frac{h\nu^{3+\beta}\rho(\nu)}{c^2}\frac{1}{e^{\frac{h\nu}{k_BT_D}}-1} \equiv A_i(\hat p) f_{0\beta}(\nu)
\end{align}
for $i \in \{I,Q,U\}$. Here the prefactor $A_i(\hat p)$ encodes the position-dependent
dust opacity and polarization.
Since $T$ and $\beta$ are constant, the frequency-dependent part of this spectrum is
already position-independent, so we don't actually need to Taylor-expand in this case.
We just need to precompute a single autocorrelation shape which is rescaled for each
pointing.
\begin{align}
I^\textrm{dust}_{\Delta t,i}(\hat p,\Delta t) &= A_i(\hat p) \bar f_{0\beta}(\Delta t).
\end{align}
This will need to be modified for more complicated dust models. If $T_D$ or $\beta$
only change slightly, then the Taylor expansion approach can be used. For more
substantial variation, a better approach may be to model it as several dust components,
each with fixed parameters.

The input dust map was simulated using the Python Sky Model \cite{pysm-2017} code \code{PySM}
of a thermal dust-only sky evaluated at 600 GHz (with no bandpass).
This was computed at HEALPix $N_\textrm{side} = 512$, but the polarization
map \code{PySM} produces is limited to $2\degree$ resolution due to the limited
resolution of the Planck polarized dust maps it uses as input. This HEALPix map
was then repixelized to $0.1\degree$ equirectangular (CAR) pixelization in
ecliptic coordinates by computing its spherical harmonics coefficients,
projecting these onto CAR, and then rotating from galactic to ecliptic coordinates
using bicubic spline interpolation and rotating the polarization vectors to compensate.

\subsubsection{Other components}
The results reported here are based on simulations that only include CMB and dust,
but other components such as synchrotron, free-free, CO, AME, etc. can be implemented
in a similar vein as above, as long as they can be approximated as a sum of
constant-spectral-shape components or can be Taylor-expanded to sufficient accuracy.

\section{Beams and sidelobes}
PIXIE will use multi-moded optics, as opposed to the single-moded optics common to
many imaging CMB instruments. For a single-moded system, the
number of modes is fixed at unity and the beam size depends on the
etendue and observing wavelength ($\Omega \propto \lambda^2$).
For a multi-moded system, the beam size is fixed and the number of modes
depends on the etendue and wavelength ($N_\textrm{modes} \propto \Omega / \lambda^2$).
This has been tested in the lab for PIXIE, and the beam was found to be
\emph{frequency-independent} above about 30 GHz \citep{pixie-multi-moded-2015,pixie-multi-moded-beams-2018}.
PIXIE's beam
is also approximately top-hat shaped, but we will here approximate it with
a Gaussian with FWHM of $1.9\degree$.

As discussed in section~\ref{sect:autocorr-problems},
a frequency-independent beam is much cheaper to implement than a frequency-dependent
one as long as the spectrum maps are linear functions of a small number of input
maps. For our dust model this is simple - the spectrum is proportional
to a single spatially varying dust opacity map, so it is sufficient to apply the
beam to that map.

\subsection{A small error in the CMB beam treatment}
The CMB, on the other hand, is modeled as a 4th order Taylor
expansion in $\Delta T$, so in this case we should smooth each power of $\Delta T$
individually. We currently \emph{do not do this}. Instead, we simply smooth
the input $\Delta T$, as one usually does when simulating CMB maps. This reduces
the accuracy of our Taylor expansion, and should be fixed in a future release.
However, this is not as serious as one might fear.
\begin{enumerate}
	\item The main reason why we go to 4th order is the $O(10^{-3})$ CMB dipole,
		but the dipole is practically unaffected by the beam. The beam-relevant
		scales are much lower, at $O(10^{-5})$. The first incorrect correction
		term is the second order, which is down by another such factor, giving
		a relative accuracy of $10^{-5}$ for T, E and B perturbations. This is
		dwarfed by cosmic variance and noise for all PIXIE scales.
	\item The same beam incorrect smoothing is used when evaluating the accuracy of the
		recovered maps. These errors therefore cancel, and the difference maps and
		error plots in the results section are identical to those we would have gotten
		if this error had not been made.
\end{enumerate}

\subsection{Sidelobes and asymmetric beams}
So far we've assumed that the beam is isotropic and position-independent,
so it can be implemented by a one-time smoothing of the maps. A full
implementation of general beam shapes would be very expensive, as it
requires an integral over (part of) the sky for every sample generated.
However, we can capture all the interesting effects of complicated beams
by expanding them as a series of symmetric beams with different pointing
offsets. For example, a slightly elliptical beam can be approximated as
the sum of two slightly offset symmetric beams.
We implement this by replacing every evaluation of the sky autocorrelation
function with a sum over such evaluations for each beam component.
Each such beam component is defined by specifying
$\Delta\phi$ and $\Delta\theta$ (see eq.~(\ref{eq:barrel})),
a beam profile, and a Muller matrix which encodes its
intensity and leakage properties.

\subsection{Simulator pseudo-python}
The full source code of the simulator can be found in the
classes \texttt{OpticsSim} and \texttt{ReadoutSim} in
\texttt{pixie.py} in \url{https://github.com/amaurea/pixie}, but the
overall logic is summarized in the pseudo-code below.
\begin{lstlisting}
for each barrel, beam:
	skies[barrel.sky, beam] = prepare_sky(barrel.sky, beam)
# Time-ordered data is generated and output in units of scans
for each scan:
	tod = zeros([ndet,nsamp*nsub])
	for each subsample in scan:
		elements = calc_orbital_parameters(subsample)
		for each detector:
			for each beam seen through each barrel by detector:
				p   = calc_pointing(elements, beam)
				sky = skies[barrel.sky, beam]
				# Compute the autocorrelation function at both
				# dt=0 (DC) and dt=p.delay (offset)
				sky_signal   = calc_sky_autocorr(sky, p)
				# Compute detector response to DC and offset sky Stokes parameters
				det_response = calc_det_response(det, barrel, p)
				tod[det,subsample] = calc_det_signal(sky_signal, det_response)
	# This uses Gaussian quadrature to integrate the subsamples
	# in each sample (so the subsamples are not equi-spaced)
	tod = downsample(tod, nsub)
	output(tod)
\end{lstlisting}

\section{PIXIE map-making}
\subsection{Assumptions}
PIXIE's observing strategy is designed to make mapmaking
fast and accurate. To be as general as possible we avoided depending
on these features in the TOD simulator, but we make use of them
in the map-maker.

\paragraph{No DC signal}
The unmodulated component (DC) of the signal changes on minute timescales,
which will not be recoverable due to correlated noise and the TOD highpass
filter. We ignore this component in the map-maker, and simply treat it as
a part of the noise.

\paragraph{Single sky}
We assume that the calibrator in single barrel mode is constant and
unpolarized. That means that its effect can be taken into account
simply by adding back the calibrator spectrum in map space, and
the map-maker itself can ignore it, and set $\vec s^\textrm{sky}_B=0$.
We also assume that the sky itself is time-independent, so there is just
a single sky to solve for. Any deviations from this will be interpreted as noise.

In double barrel mode we assume that the two barrels see the same signal,
and that the V and U antisymmetric leakage terms can be ignored.
Since I can't be recovered in double barrel mode, the two barrels are
equivalent to a single barrel with no I but twice the polarization signal,
so we can again set $\vec s^\textrm{sky}_B=0$ as long as we multiply the
detector response by 2. With this, equation~(\ref{eq:response}) simplifies
to

\begin{align}
	\vec s^\textrm{det}(t) &=
		G \cdot R_\textrm{spin}(t)\cdot\vec s^\textrm{sky}(\hat p_t,\Delta t)
\end{align}
where $G$ is the detector-barrel response matrix
$G = \frac{g}{4}[\vec e_I+\vec e_Q,\vec e_I-\vec e_Q,-\vec e_I+\vec e_Q,
-\vec e_I-\vec e_Q]$ and $g = 1$ in single barrel mode and $g = 2$ in
double barrel mode.

\paragraph{Regular scanning pattern}
PIXIE scans in great circles (rings) that pass through the ecliptic poles,
and both barrels are perfectly aligned. This means that the scanning motion
does not induce polarization rotation in ecliptic coordinates, and that all
samples in a ring will have constant ecliptic longitude, up to pole wrapping.
The sampling rate and telescope scan, spin and stroke speed are constant;
and there is an integer number of samples per stroke, strokes per spin and spins
per scan. Using this, the data model simplifies further to

\begin{align}
	\vec s^\textrm{sky}_{ri} &= G \cdot
		R_\textrm{stok}\Big(\frac{4\pi i}{N_\textrm{spin}}\Big) \cdot
		\vec s^\textrm{sky}\Big(\Big[l=l_0+r\Delta l,b=b_0+\frac{2\pi i}{N_\textrm{scan}}\Big],
		A_\textrm{delay}\textrm{triangle}\Big(\frac{i}{N_\textrm{stroke}}\Big)\Big)
		\label{eq:sring}
\end{align}
where $r$ is the ring index and $i$ is the sample inside the ring (such
that the total sample number in the time-ordered data is $rN_\textrm{scan}+i$),
where $N_\textrm{stroke} = 1\,920$, $N_\textrm{spin} = P_\textrm{spin}N_\textrm{stroke}
= 15\,360$ and $N_\textrm{scan} = P_\textrm{scan}N_\textrm{spin} = 5\,898\,240$
are the number of samples per scan, spin and stroke respectively, and where
$P_\textrm{spin} = 8$ and $P_\textrm{scan} = 384$ are the number of
strokes per spin and spins per scan respectively.

\paragraph{Simple noise}
For simplicity, we will also assume that the detectors have independent white
noise of equal amplitude. Neither of these are necessary, and can be easily
relaxed in the future. Section~\ref{sect:corrnoise} discusses the effect
of correlated (1/f) noise on the PIXIE spectral maps.

With the nominal rates of 1024 detector samples per mirror stroke,
the Fourier transform returns 512 synthesized frequency channels each
14.4 GHz wide, extending from DC to 7.4 THz. Scattering filters
in the optical path limit the response at frequencies above 6 THz.
Dispersion within the FTS from the finite spread of angles within
the beam washes out the fringe amplitude at comparable frequencies.
Synthesized channels at frequencies above $\sim$6 THz thus contain
noise but no signal, providing a convenient check of the instrument
noise.

\begin{figure}
	\centering
	\includegraphics[width=70mm]{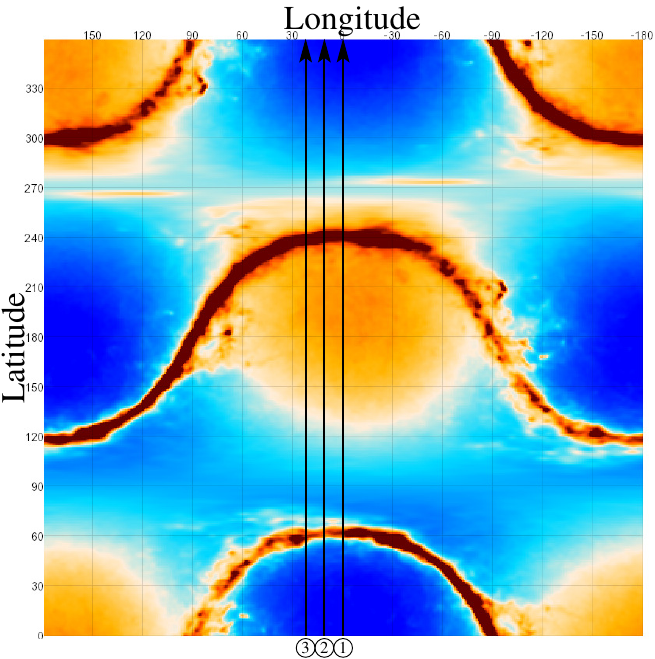}
	\caption{PIXIE's
	scanning pattern is particularly simple in ecliptic (extended) plate carrée (CAR)
	coordinates, where each scan moves at constant pixel velocity purely along the
	latidude axis. This allows an exact correspondence between samples and pixels.
	3 example scans separated by 7.6 days are as black lines.
	The extended CAR coordinates here cover two mirror images of the full sky
	in order to show full scans without breaks.}
\end{figure}

\subsection{Orthogonalization}
\label{sect:ortho}
\begin{figure}
	\centering
	\begin{tabular}{ccc}
		\includegraphics[width=0.32\textwidth]{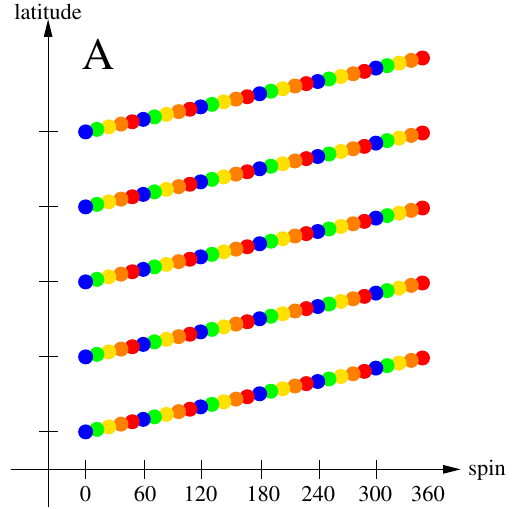} &
		\includegraphics[width=0.32\textwidth]{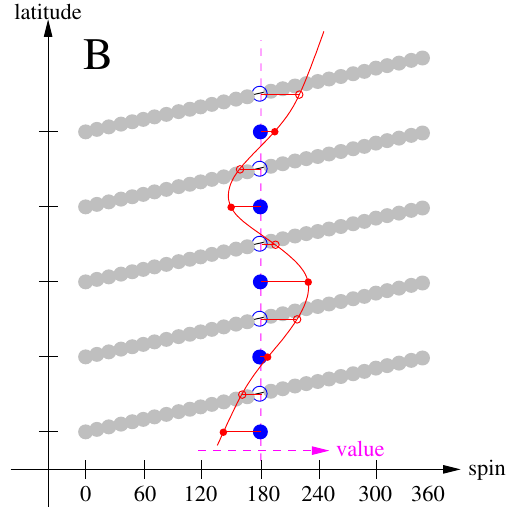} &
		\includegraphics[width=0.32\textwidth]{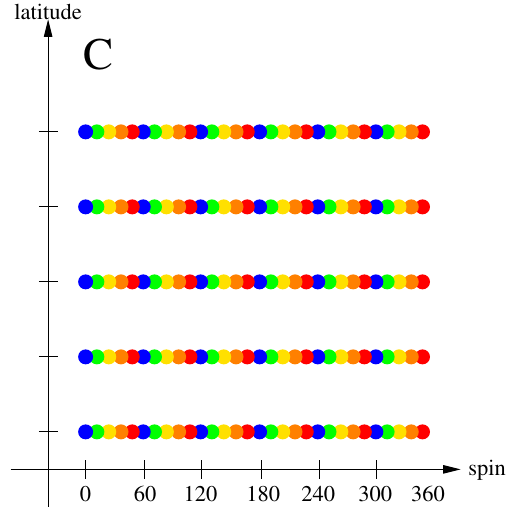} \\
		\includegraphics[width=0.32\textwidth]{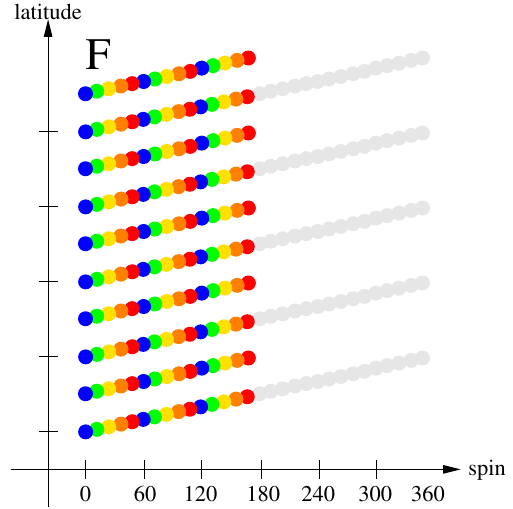} &
		\includegraphics[width=0.32\textwidth]{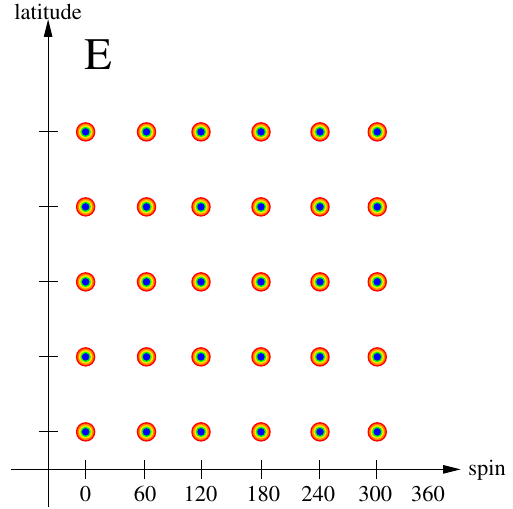} &
		\includegraphics[width=0.32\textwidth]{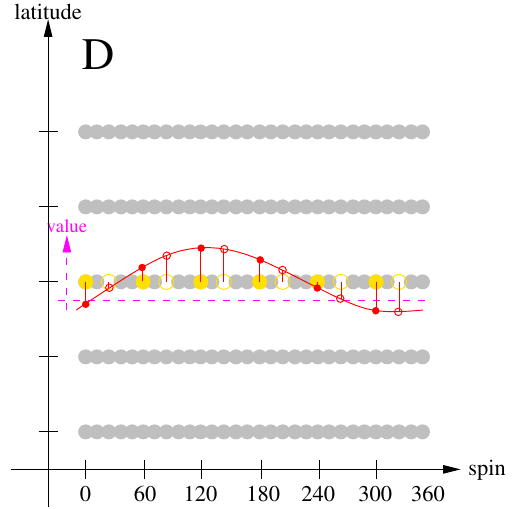}
	\end{tabular}
	\caption{Orthogonalization of the scan, spin and stroke
	motions. PIXIE scans in circles on the sky, so the
	signal is periodic during a scan. It also spins and strokes.
	In isolation the signal would also be a periodic function
	of the spin angle and mirror stroke position, but because
	PIXIE scans while it spins, and spins
	while the mirror strokes these dimensions get mixed and the
	periodicity is lost on all but the scan timescale.
	\textbf{A}: Illustration of the mixing of the stroke (color),
	spin (x-axis) and scan (y-axis) dimensions in the PIXIE scanning
	pattern. \textbf{B}: A subset of points with the same stroke
	position and spin angle from periodic functions of the scan
	angle. The open circles represent the positions at which
	we have a measurement, while the filled ones are the positions
	we want to estimate. The red curve represents the smoothly
	varying signal inferred from the open points, which is used to
	interpolate the values at the filled points. The effect of
	this is to undo the scanning motion during each spin.
	Note that the red curve has its own x axis corresponding
	to the time-ordered data values.
	\textbf{C}: Spin and stroke are still
	mixed. \textbf{D}: We Fourier-interpolate between points with
	the same stroke position within a spin to undo the spin
	motion during each stroke. \textbf{E}: The scan, spin and stroke
	are now unmixed. The resulting timestream simulates what we would
	see if PIXIE stood still while the mirror strokes, then
	instantanously rotates to a new spin angle, strokes again, etc.
	and then instantanously moves to a new scan angle when the spin is done.
	\textbf{F}: The signal is spin-0 (I) or spin-2 (Q,U), so we can
	shorten the interpolation distance and decrease the pixel size
	by mapping the $180\degree-360\degree$ interval to the $0\degree-
	180\degree$ interval in spin angle.}
	\label{fig:ortho}
\end{figure}
It would have been very convenient if the stroke, spin and scan motions were
orthogonal, so that the telescope didn't spin during a stroke and didn't scan
during a spin. That would allow us to demodulate each of them independently.
Rather than have stroke, spin and pointing change smoothly during each scan
we would have $P_\textrm{scan}$ pointings (corresponding
to our output pixels), each with
$P_\textrm{spin}$ spin angles measured, and for each
of those $N_\textrm{delay}$ mirror positions.

Figure~\ref{fig:ortho} illustrates the situation. In the 3-dimensional
space of latitude, spin angle and mirror position (corresponding to the
horizontal, vertical and color axis in the figure), the scanning pattern
traces out a skewed grid (A) instead of the orthogonal grid
that would be convenient (E). However, as long as our signal is well-sampled
in all directions in the grid we can use the samples we have to
interpolate to the orthogonal grid we wish we had.

We start by undoing the effect of the scan motion during each scan (and stroke)
by splitting our samples into groups with the same spin angle and mirror position.
Letting $d_i$ be the i'th sample of the time-ordered data $d$ for a single detector
for a single ring, we split into groups such that
sample (pixel) $p$ in group $g \in \{0,N_\textrm{spin}-1\}$ corresponds to
$i = pN_\textrm{spin}+g$, and let the notation $d_{ps} \equiv d_{i=pN_\textrm{spin}+g}$.
Because the shorter time-scale parameters are constant within each group, they see
a signal that changes smoothly as the telescope scans over the beam-smoothed sky
(i.e. as $p$ changes while $g$ is fixed),
and this makes it easy to interpolate. Since the ring time-ordered data as a whole
is periodic each of these evenly sampled ring subsets is also periodic, so we
can interpolate via Fourier shifting: Given an array $a$ containing $N$ samples and a discrete Fourier transform $F$,
an array $a'$ equal to $a$ shifted downwards by $\Delta$ samples is given by
\begin{align}
	a_j' = a_{j+\Delta} &= \Psi(\Delta)_{jk} a_k \equiv F^{-1}_{jf} e^{-2\pi if\Delta/N} F_{fk} a_k
\end{align}
which defines the Fourier-shift matrix $\Psi(\Delta)$. Using this,
the scan drift corrected tod $d'$ is given by:
\begin{align}
	\vec s^\textrm{shift}_{rpg} &= \Psi\big(g/N_\textrm{spin}\big)_{pp'} \vec s^\textrm{det}_{rp'g}
\end{align}
corresponding to panel C in the figure. This involves interpolating
a distance of up to $\frac12\cdot 360\degree\cdot N_\textrm{spin}/N_\textrm{scan}
= 0.46875\degree$ on the sky, of about half a PIXIE FWHM. We can halve the
interpolation distance by exploiting the spin-2 nature of our signal:
only half a spin rotation is needed to return to the same configuration (panel F).
This spin-2 scan drift correction is identical to the one above, but with
half the normal value of $N_\textrm{spin}$.

Undoing the effect of PIXIE's spin during the mirror motion follows the same
logic on shorter time scales. We now group by mirror position (panel D)
such that
\begin{align}
	\vec s^\textrm{ort}_{rs\delta} &= \Psi\big(\delta/N_\textrm{stroke}\big)_{ss'} \vec s^\textrm{shift}_{rs'\delta}
\end{align}
with $i = sN_\textrm{stroke}+\delta$, $s$ being the number of completed strokes
and $\delta$ being the index in the mirror displacement pattern.

The orthogonalized equivalent to eq.~\ref{eq:sring} is
\begin{align}
	\vec s^\textrm{ort}_{rps\delta} &= G R_\textrm{stok}
		\Big(\frac{4\pi s}{P_\textrm{spin}}\Big) \vec s^\textrm{sky}_{rp\delta} &
		\textrm{with} & &
	\vec s^\textrm{sky}_{rp\delta} &=
		\vec s^\textrm{sky}\Big([l_r,b_p],
		A_\textrm{delay}\textrm{triangle}\Big(\frac{\delta}{N_\textrm{stroke}}\Big)\Big)
	\label{eq:sort}
\end{align}
and where $l_r = l_0 + r\Delta l$ is the longitude of ring $r$,
$b_p = b_0 + \frac{2\pi p}{P_\textrm{scan}}$ is the latitude of pixel $p$
along the ring.

\subsection{Spin demodulation}
Now that position, spin angle and mirror position are independent, we can
handle each of them separately. We start by demodulating the spin.
Suppressing the indices $r$, $p$ and $\delta$ to avoid excessive
verbosity, and expanding $\vec s^\textrm{sky} = [I,Q,U]$, 
equation~\ref{eq:sort} evaluates to
\begin{align}
	s^\textrm{ort}_{ds} &= \left[G_{dI} I +
		(G_{dQ}-iG_{dU})e^{4\pi i s/P_\textrm{spin}} (Q + iU)\right]
\end{align}
This is just a weighted sum of 3 orthogonal Fourier modes,
and can be straightforwardly inverted as
\begin{align}
	I &= \langle G_{dI}^{-1} s^\textrm{ort}_{ds}\rangle_{ds} &
	Q &= \Re \langle G_{dQ}^{-1}
			e^{-4\pi i s/P_\textrm{spin}} s^\textrm{ort}_{ds}\rangle_{ds} &
	U &= \Im \langle G_{dQ}^{-1}
			e^{-4\pi i s/P_\textrm{spin}} s^\textrm{ort}_{ds}\rangle_{ds}
\end{align}
which is simply the gain-correct 0th Fourier mode and
real and imaginary part of the 2nd Fourier mode.

\subsection{Frequency response correction}
PIXIE's response drops off as we approach 6 GHz (see section~\ref{sect:response}).
We correct for this by dividing each frequency bin by the instrument frequency
response function $\rho(\nu)$ evaluated at the bin center.

\subsection{Mapmaker pseudo-python}
\label{sect:mapmaker}
The full source code of the map-maker can be found in the
programs \texttt{tod2ring.py} and \texttt{ring2map.py} in
\url{https://github.com/amaurea/pixie}, but the overall logic is summarized in
the pseudo-code below.
\begin{lstlisting}
map  = zeros([nfreq,{I,Q,U},nlat,nlon])
hits = zeros([nlat,nlon])
for each scan:
	d = unapply_tod_filter(scan.tod)
	d = unapply_sample_window(d)
	# Unapply scan drift during spin
	if spin == 1:
		d = d.reshape(ndet,scanspins,spinstrokes*strokesamps)
		d = fourier_shift(d, range(d.shape[-1])/d.shape[-1], axis=1)
	else: # spin-2 - could be modified to double resolution
		d = d.reshape(ndet,scanspins*2,spinstrokes*strokesamps/2)
		dhalf = fourier_shift(d, range(d.shape[-1])/d.shape[-1], axis=1)
		d[:,0::2], d[:,1::2] = dhalf, dhalf
	# Unapply scan and spin drift during stroke
	d = d.reshape(ndet,scanspins*spinstrokes,strokesamps)
	d = fourier_shift(d, range(d.shape[-1])/d.shape[-1], axis=1)
	# Decompose spin modulation into I (spin-0 part) and Q,U (spin-2 part)
	d = d.reshape(ndet,scanspins,spinstrokes,strokesamps)
	d = fft(d, axis=2)
	# Transform from autocorrelation into spectrum
	d = [d[:,:,0],d[:,:,2].real*2,d[:,:,2].imag*2]
	d = fft(d, axis=3)*2/strokesamps/dfreq
	# Take into account the detector polarization orientation and frequency response
	d = unapply_detector_response(d)
	add_to_sky(map,  scan.lon, d)
	add_to_sky(hits, scan.lon, 1)
map /= hits
output(map)
\end{lstlisting}

\section{Performance}
The software was tested on the Scinet GPC cluster. On a typical node
with an 8-core Intel Xeon 2.5 GHz processor the simulator has a run time of
about $T_\textrm{sim} = 81 \textrm{ sec} \cdot N_\textrm{scan} N_\textrm{sub} / N_\textrm{core}$,
and the map-maker has $T_\textrm{map} = 23 \textrm{ sec} \cdot N_\textrm{scan} / N_\textrm{core}$.
A 192-scan run with 9 subsamples per sample, like the ones used in this article
therefore takes about 40 CPU-hours. The full PIXIE scanning pattern, with 1369.7 scans
per year, would take 286 CPU-hours per simulated year.

The large speed difference between the simulator and map-maker is partially due to
the overhead from high spatial and spectral resolution simulations, which are necessary
for investigating sub-pixel and sub-sample biases. If one is not interested in
these, then disabling pixel window integration results in a factor $\sim 7$ speedup.
Further speedups would be possible in a simulator that operates at the same spatial
and spectral resolution as the final maps, such as the spectral cube design needed
for spatially and spectrally varying beams (see section~\ref{sect:autocorr-problems}).

Since each scan takes 384 minutes to collect but only 23 CPU-seconds to map with this
mapmaker, a single core would easily keep up with the data down-link from the instrument.

\begin{figure}
	\centering
	\hspace*{-15mm}\begin{tabular}{rm{56mm}m{54.4mm}m{56mm}}
		& \hspace{32mm}I & \hspace{32mm}Q & \hspace{32mm}U \\
		CMB &
		\includegraphics[height=28mm,clip,trim=0 8mm 7.5mm 0]{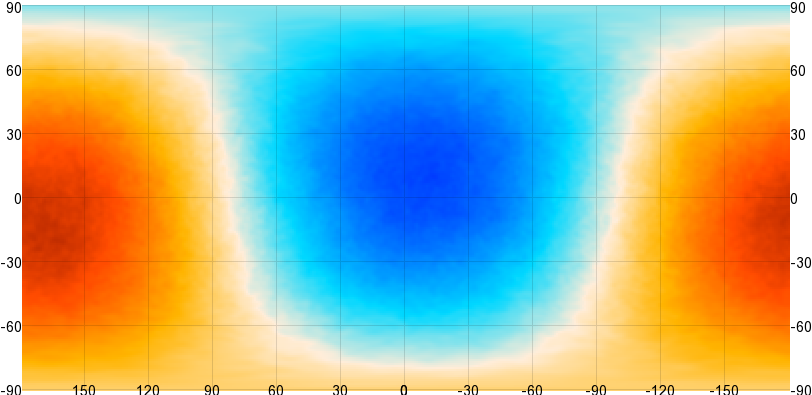} &
		\includegraphics[height=28mm,clip,trim=7.5mm 8mm 7.5mm 0]{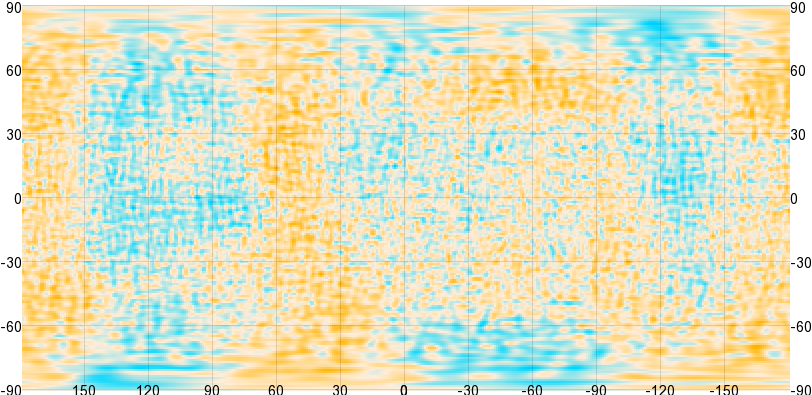} &
		\includegraphics[height=28mm,clip,trim=7.5mm 8mm 0 0]{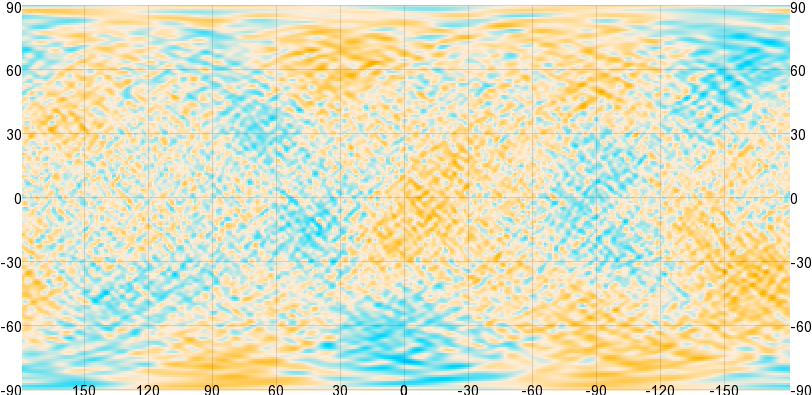} \\
		Dust &
		\includegraphics[height=28mm,clip,trim=0 8mm 7.5mm 0]{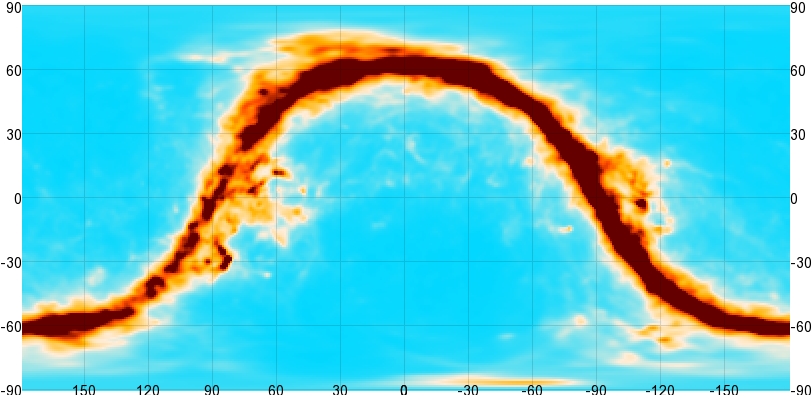} &
		\includegraphics[height=28mm,clip,trim=7.5mm 8mm 7.5mm 0]{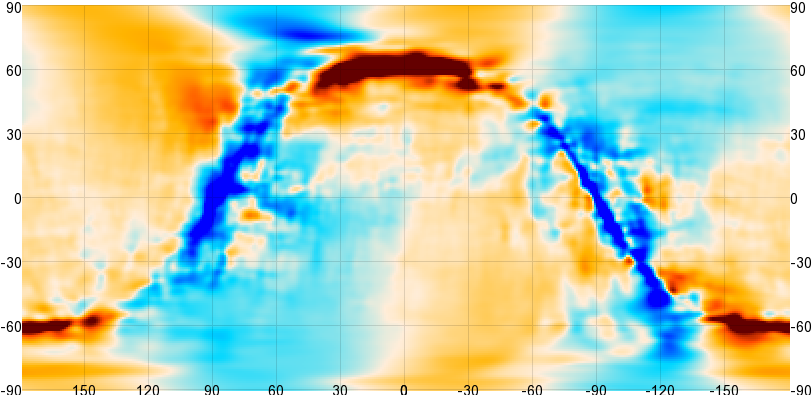} &
		\includegraphics[height=28mm,clip,trim=7.5mm 8mm 0 0]{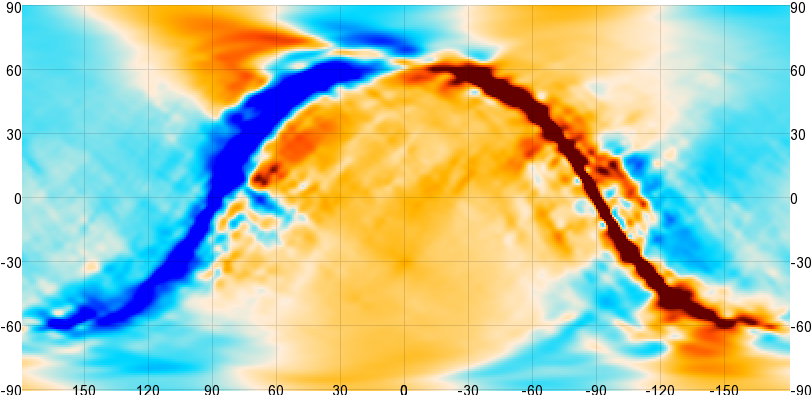} \\
		Total &
		\includegraphics[height=29.7mm,clip,trim=0 0mm 7.5mm 0]{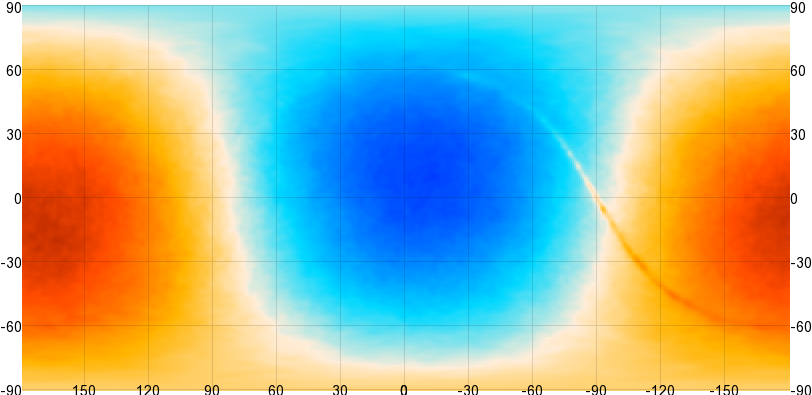} &
		\includegraphics[height=29.7mm,clip,trim=7.5mm 0mm 7.5mm 0]{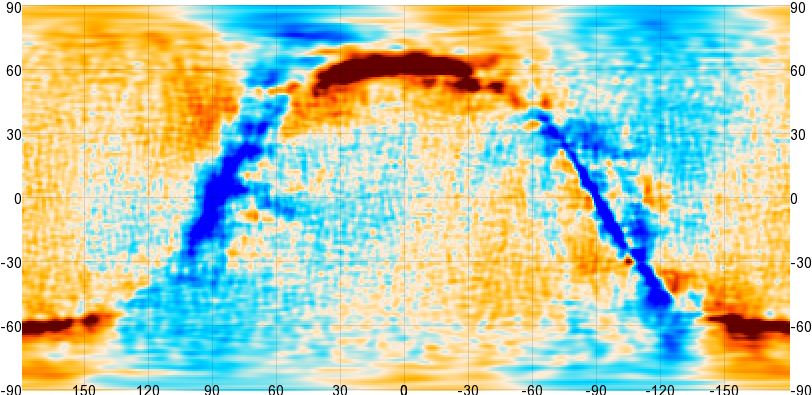} &
		\includegraphics[height=29.7mm,clip,trim=7.5mm 0mm 0 0]{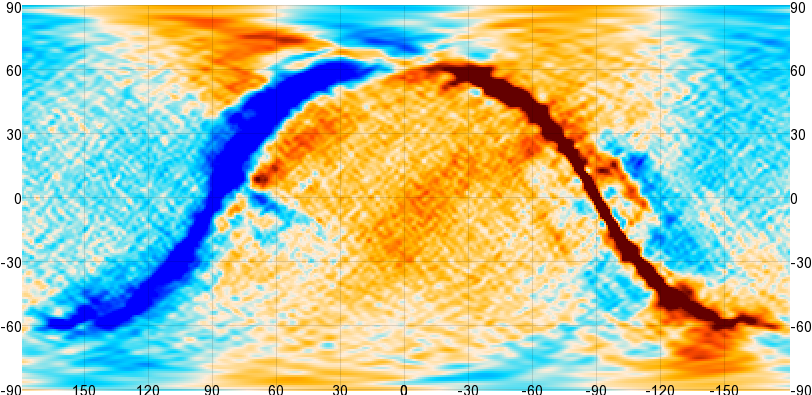}
	\end{tabular}
	\caption{The input beam-smoothed sky model evaluated at 58 GHz in the CAR
	projection in ecliptic coordinates. The color range is
	$\pm 400$ kJy/sr in intentisty (I) and $\pm 200$ Jy/sr in polarization (P), except for dust I where it is
	$\pm 5$ kJy/sr. The monopole has been subtracted in each map for plotting purposes.}
	\label{fig:model}
\end{figure}

\section{Results}
\label{sect:results}
We generated a simple two-component CMB + dust sky model as described in
section~\ref{sect:cmb}-\ref{sect:dust}. A 58 GHz slice of this
model can be seen in figure~\ref{fig:model}. We then simulated
scanning across this model with the accelerated scanning pattern
described in section~\ref{sec:pointing}, which would result in 384
scans per orbit.
However, since the scans are great circles, only half an orbit
is needed to cover the whole sky. Our simulations therefore consist
of 192 scans, each consisting of 5898240 samples for each of 4
detectors at 256 Hz.

\begin{figure}
	\centering
	\begin{tabular}{cc}
		\includegraphics[width=0.48\textwidth,trim=8mm 0 0mm 0mm]{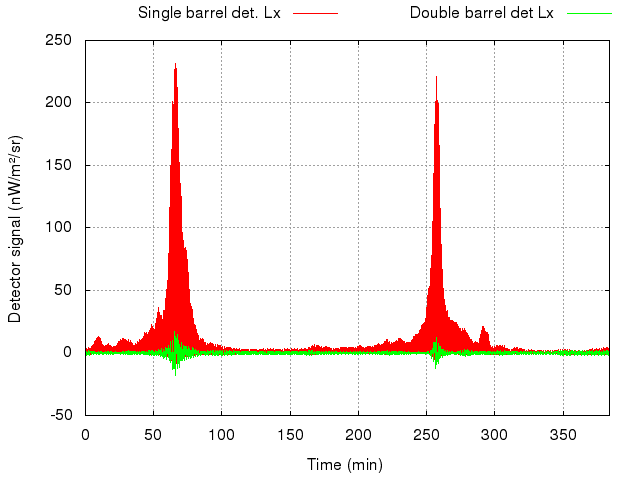} &
		\includegraphics[width=0.48\textwidth,trim=8mm 0 0mm 0mm]{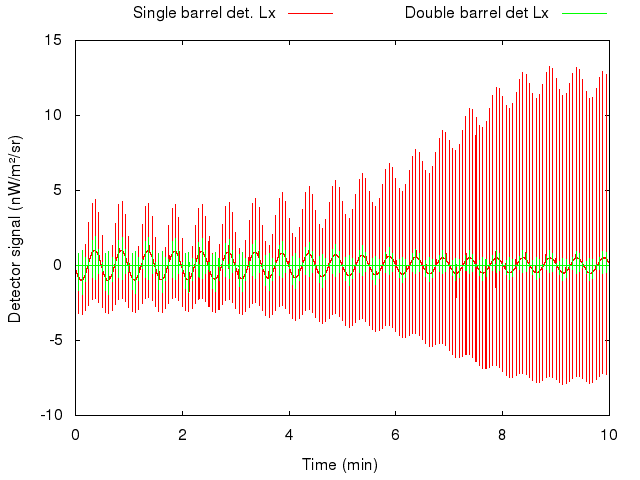} \\
		\includegraphics[width=0.48\textwidth,trim=8mm 0 0mm 0mm]{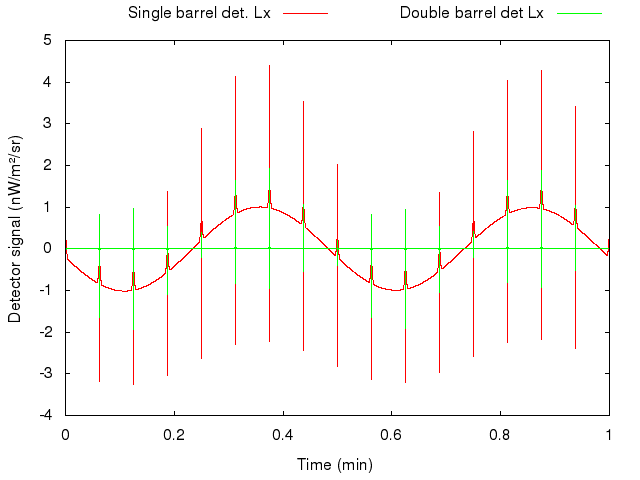} &
		\includegraphics[width=0.48\textwidth,trim=8mm 0 0mm 0mm]{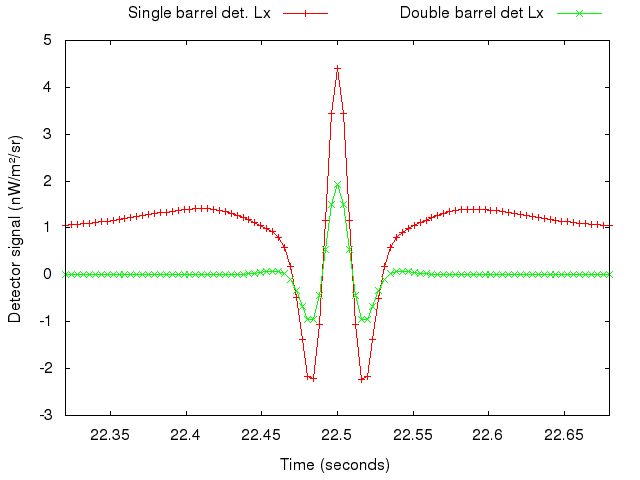}
	\end{tabular}
	\caption{Noiseless simulated PIXIE time-ordered data for the Lx
	detector on various time scales in single (red) and double (green)
	barrel observing mode. \emph{Top left}: A whole great-circle
	scan of the sky, starting at ecliptic coordinates $l=0\degree$,
	$b=0\degree$ and scanning in longitude. The two peaks are crossings
	of the galactic plane. \emph{Top right}: Zoom on the first 10 minutes.
	We see that the signal is modulated on 3 time-scales: The sky signal
	changes on minute time-scales due to the scan; the polarization is
	modulated on 15 second time-scales due to PIXIE's spin; and the
	signal is modulated on second time-scales by the mirror stroke.
	\emph{Bottom left}: Zoom on a single PIXIE spin. The slowly
	changing baseline is the DC signal, which PIXIE will not attempt
	to measure due to its susceptibility to 1/f noise. \emph{Bottom
	right}: Zoom on a single mirror half-stroke, centered on
	$\Delta t=0$. This is effectively a plot of the electric field's
	autocorrelation function at this position.}
	\label{fig:tod}
\end{figure}

Figure~\ref{fig:tod} shows an example of what the time-ordered
data looks like on various time scales, here for a noiseless
single detector, in both single and double barrel mode. The
stroke and spin modulation makes the TOD quite different from
that of standard CMB experiments, which have a smoothly varying
signal on short time scales.

\begin{figure}
	\centering
	\hspace*{-15mm}\begin{tabular}{rm{56mm}m{54.4mm}m{56mm}}
		& \hspace{32mm}I & \hspace{32mm}Q & \hspace{32mm}U \\
		Input &
		\includegraphics[height=28mm,clip,trim=0 8mm 7.5mm 0]{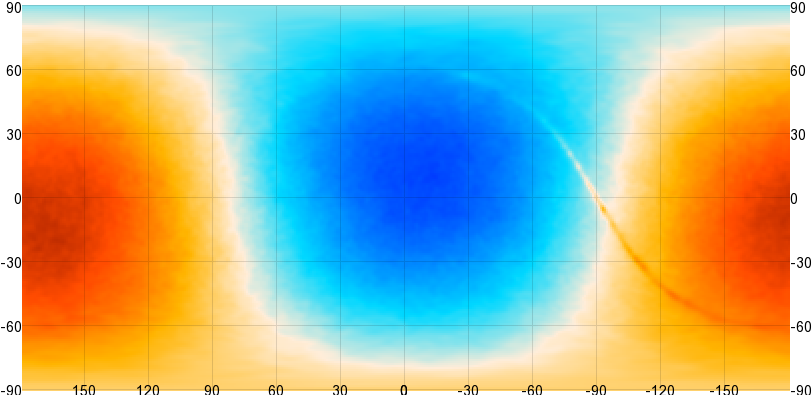} &
		\includegraphics[height=28mm,clip,trim=7.5mm 8mm 7.5mm 0]{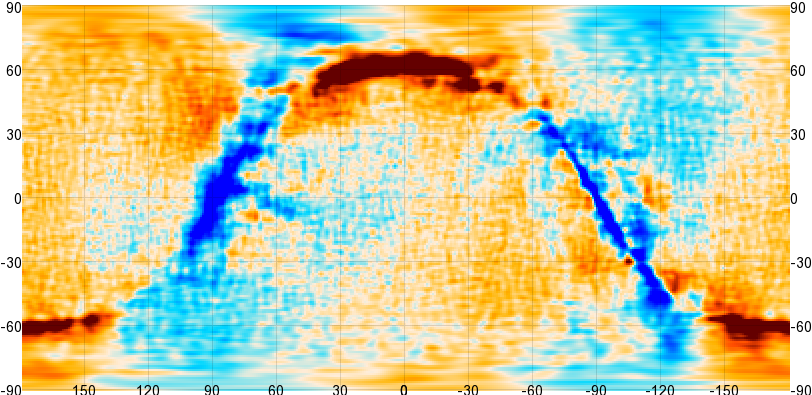} &
		\includegraphics[height=28mm,clip,trim=7.5mm 8mm 0 0]{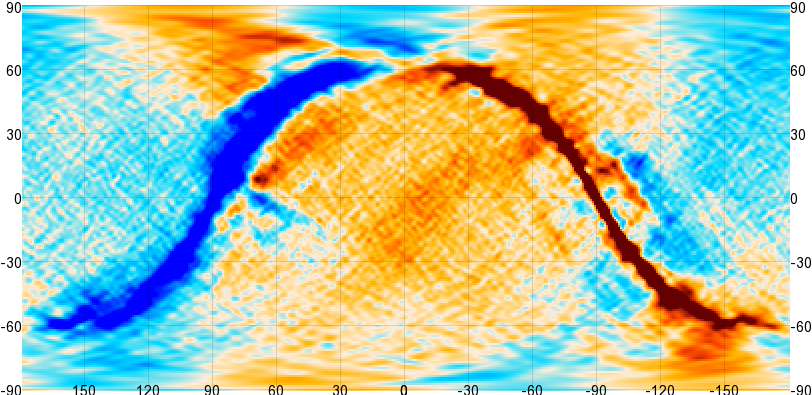} \\
		Output &
		\includegraphics[height=28mm,clip,trim=0 8mm 7.5mm 0]{plots/std_sb_map_0.png} &
		\includegraphics[height=28mm,clip,trim=7.5mm 8mm 7.5mm 0]{plots/std_sb_map_1.png} &
		\includegraphics[height=28mm,clip,trim=7.5mm 8mm 0 0]{plots/std_sb_map_2.png} \\
		Error &
		\includegraphics[height=29.7mm,clip,trim=0 0 7.5mm 0]{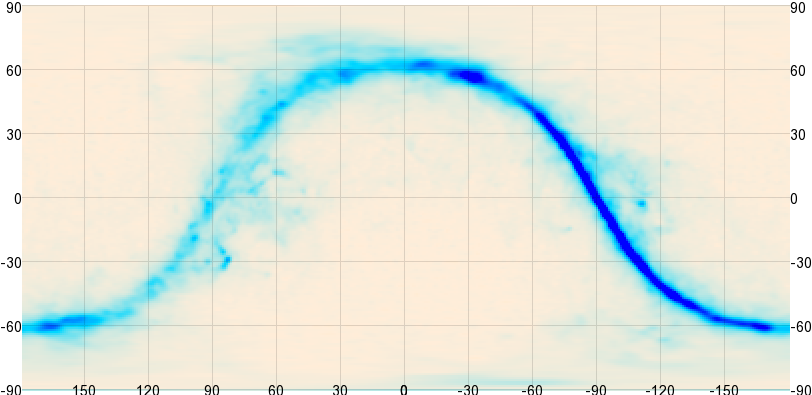} &
		\includegraphics[height=29.7mm,clip,trim=7.5mm 0 7.5mm 0]{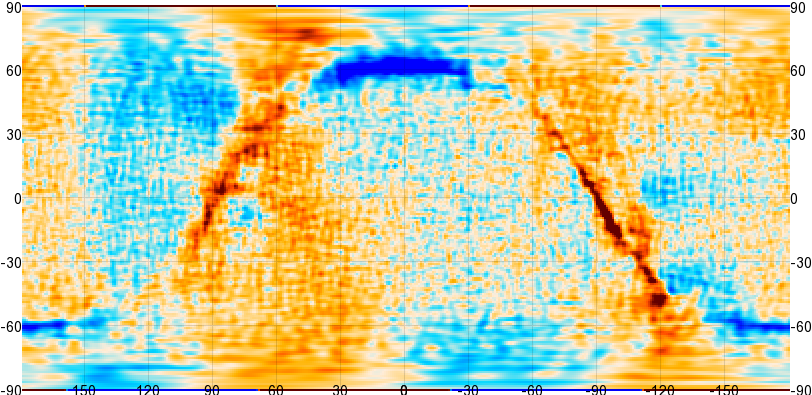} &
		\includegraphics[height=29.7mm,clip,trim=7.5mm 0 0 0]{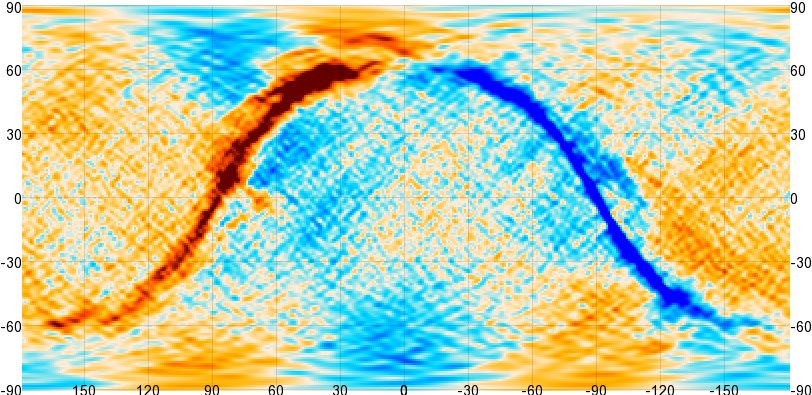}
	\end{tabular}
	\caption{Simulator/map-maker spatial bias test.
		\emph{Top row}: The beam-smoothed input map at 58 GHz.
			The color range is $\pm 400$ kJy/sr for I and $\pm 200$ Jy/sr for P.
		\emph{Middle row}: The output map at 58 GHz.
		\emph{Bottom row}: Difference between the mapmaking output maps and
			input maps at the same frequency for a noise-less simulation.
			The color range is $\pm 500$ mJy/sr for I and $\pm 5$ mJy/sr for P.
			These represent the bias of the simulator-map-maker combination.
			The biases are small compared to the instrument noise.}
	\label{fig:spatbias}
\end{figure}

\begin{figure}
	\centering
	\includegraphics[width=0.8\textwidth]{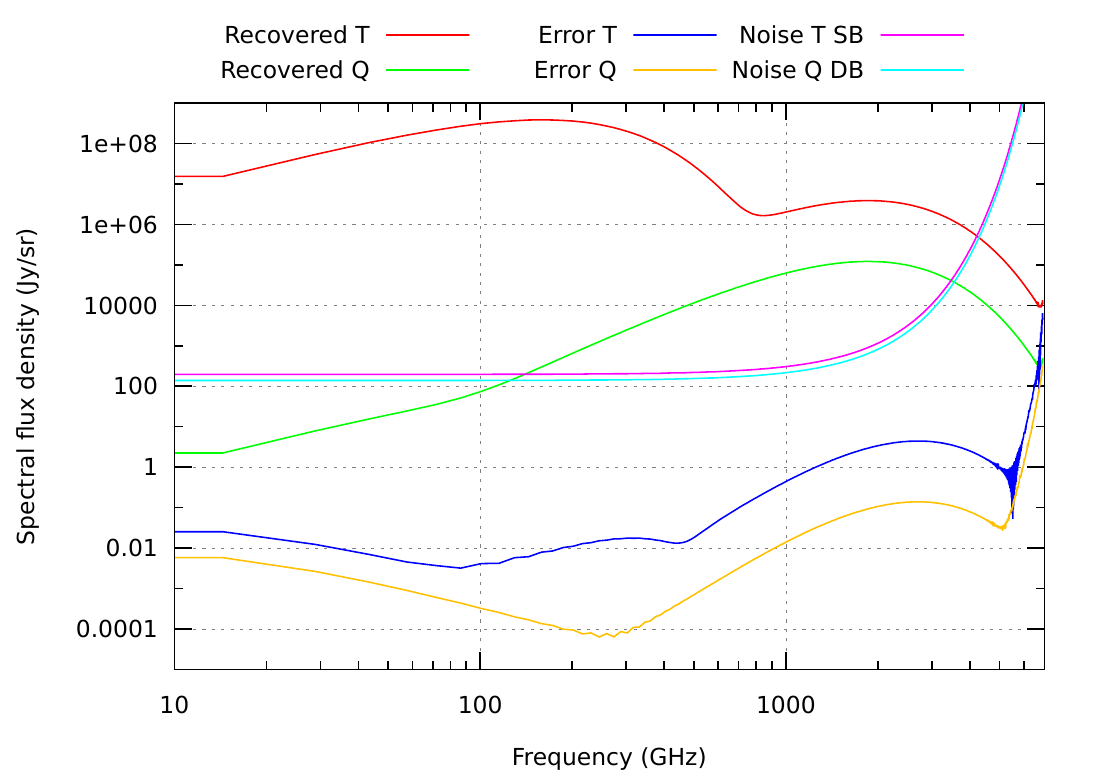}
	\caption{Simulator/map-maker spectral bias test. Recovered I and Q spectra and
	their errors for a noiseless simulation
	with 9 Gaussian quadrature subsamples per sample, compared to the PIXIE
	noise level. This is all for a single pixel at $l=0\degree$,$b=0\degree$.
	The U spectrum is similar to the Q one, but was left out to avoid clutter.
	The bias is $\sim 100$ times lower than the noise for $\nu<500$~GHz and
	$\sim 5$ times lower than the noise at the worst point at 2 THz. At 200
	GHz the bias is $\sim 10^{-10}$ of the signal in I and $\sim 10^{-6}$
	of the signal in P. These errors are independent of observing mode.}
	\label{fig:specbias}
\end{figure}

Figure~\ref{fig:spatbias} shows the
corresponding maps we get after running these simulated scans
through the mapmaker, compared to the input model at 58 GHz
evaluated at the center of each pixel. Even under these idealized
conditions the residual is not zero, though it is small: ~500 mJy/sr for
dust and much less for the CMB in T, compared to the ~100 MJy/sr
CMB monopole; and ~5 mJy/sr in P, compared to 200 Jy/sr for the signal.
This represents a $-83 \textrm{dB}$ error in T and $-46
\textrm{dB}$ in P.
Figure~\ref{fig:specbias} shows the signal and bias spectra for
a single input pixel, compared to the expected instrument noise.
These deviations are due to subsample effects, and would also
be expected when analyzing real data. See appendix~\ref{sect:subpix}
for details, but for now it suffices to note that these biases
are many orders of magnitude smaller than the instrument noise floor.

With the mapmaker successfully recovering the input signal in this ideal case,
we next study the effect of some common and less common instrument
imperfections on PIXIE’s performance.

\subsection{Intensity to polarization leakage}
\begin{figure}
	\centering
	\hspace*{-10mm}\begin{tabular}{m{35mm}rm{11mm}m{11mm}m{11mm}m{11mm}m{11mm}m{11mm}m{20mm}}
		\hspace*{7mm}Input beam &
		&
		\hspace*{5mm}Lx &
		\hspace*{5mm}Ly &
		\hspace*{2mm}Lx+Ly &
		\hspace*{2mm}Rx+Ry &
		\hspace*{5mm}All &
		\hspace*{3mm}Circ &
		\hspace*{10mm}Detail \\
		\vspace*{-11mm}\multirow{3}{*}{\includegraphics[height=43mm]{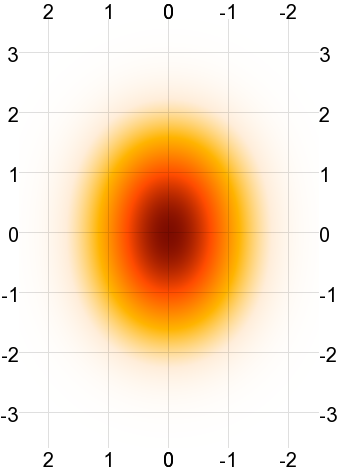}} &
		I &
		\includegraphics[height=14mm]{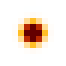} &
		\includegraphics[height=14mm]{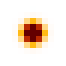} &
		\includegraphics[height=14mm]{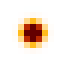} &
		\includegraphics[height=14mm]{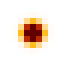} &
		\includegraphics[height=14mm]{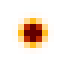} &
		\includegraphics[height=14mm]{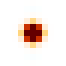} &
		\vspace*{-6mm}\multirow{3}{*}{\includegraphics[height=40mm]{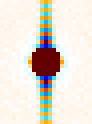}} \\
		&
		Q &
		\includegraphics[height=14mm]{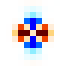} &
		\includegraphics[height=14mm]{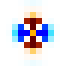} &
		\includegraphics[height=14mm]{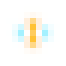} &
		\includegraphics[height=14mm]{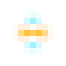} &
		\includegraphics[height=14mm]{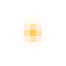} &
		\includegraphics[height=14mm]{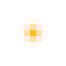} & \\
		&
		U &
		\includegraphics[height=14mm]{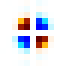} &
		\includegraphics[height=14mm]{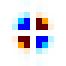} &
		\includegraphics[height=14mm]{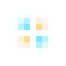} &
		\includegraphics[height=14mm]{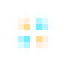} &
		\includegraphics[height=14mm]{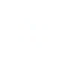} &
		\includegraphics[height=14mm]{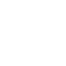} &
	\end{tabular}
	\caption{Simulation of a highly elliptical beam with a flattening of 0.5
	observing a point source with 10\% polarization in the Q direction. The
	beam is built up from 5 identical, circular, gaussian components
	with FWHM of $1.9\deg$ offset
	vertically with an interval of $0.475\deg$. \emph{Left}:
	Map of the instantaneous beam profile, with axis units of degrees. \emph{Middle}:
	The PIXIE scanning pattern circularizes the beam efficiently, at the cost
	of a slight increase in beam size compared to the non-elliptical case.
	Beam ellipticity also introduces polarization leakage due to the spin-2
	nature of an elliptical beam, but this is doubly suppressed in PIXIE.
	First, it's cancelled when combining detectors in each horn unless
	the two detectors have different beams, and secondly, it's cancelled
	when combining the signal from the two horns unless the two horns
	have different beams. In this simulation we used an unrealistically
	high
	10\% mismatch between x- and y-oriented detectors in each horn, making
	the first cancellation only 90\% efficient. The Q and U color scale is
	10\% of the I color scale here. \emph{Right}: Plotting the output beam in
	a restricted color range reveals a ringing pattern in the scan direction
	with amplitude $2.3\cdot
	10^{-3} r/1\degree$ relative to the beam peak,
	where r is the distance from beam center. This also appears
	for a circular input beam, and is a manifestation of the subsample effects
	described in section~\ref{sect:subpix}. The ringing makes it hard to
	quantify the amount of residual non-circularity, except by saying that it's
	$\lesssim 2.3\cdot 10^{-3}$ of the circular component.}
	\label{fig:src_leak}
\end{figure}

\begin{figure}
	\centering
	\hspace*{-5mm}\begin{tabular}{rm{54.4mm}m{56mm}}
		&
		%\hspace{32mm}I &
		\hspace{32mm}Q &
		\hspace{32mm}U \\
		Lx &
		\includegraphics[height=28mm,clip,trim=7.5mm 8mm 7.5mm 0]{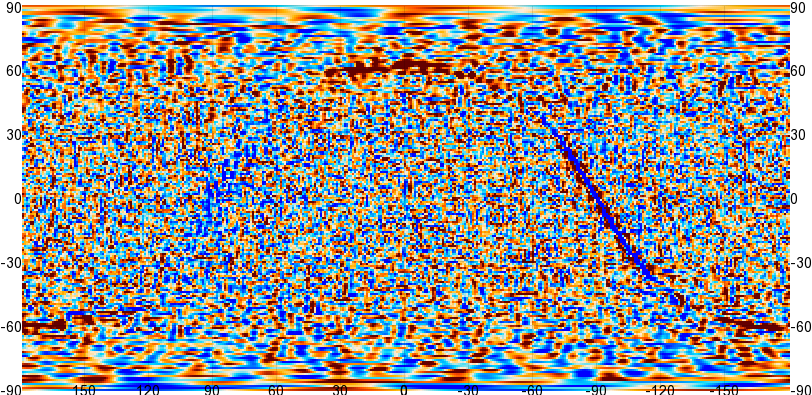} &
		\includegraphics[height=28mm,clip,trim=7.5mm 8mm 0 0]{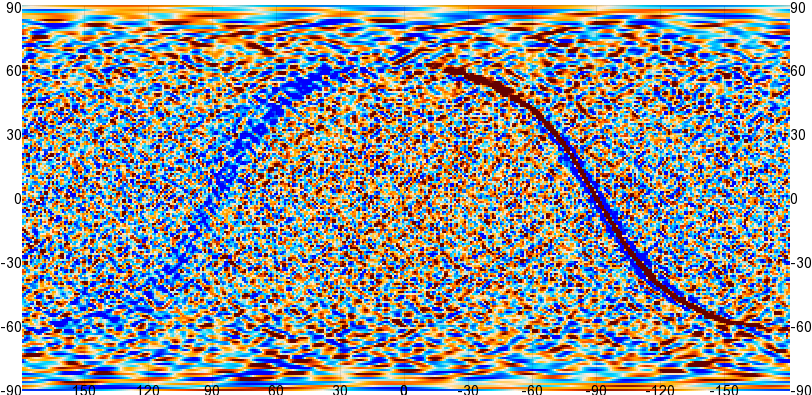} \\
		Lx+Ly &
		\includegraphics[height=28mm,clip,trim=7.5mm 8mm 7.5mm 0]{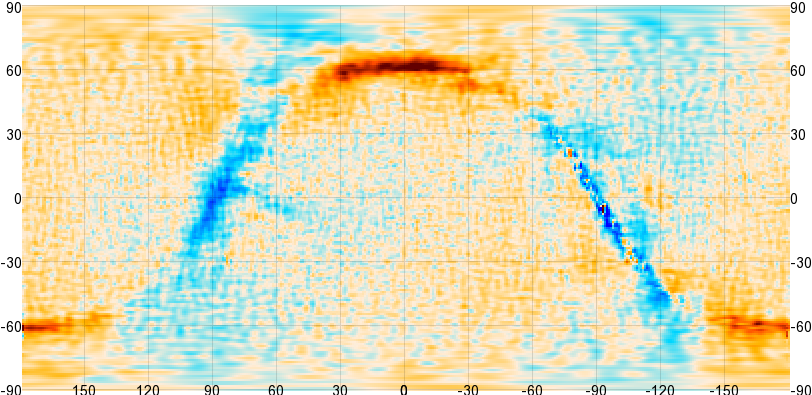} &
		\includegraphics[height=28mm,clip,trim=7.5mm 8mm 0 0]{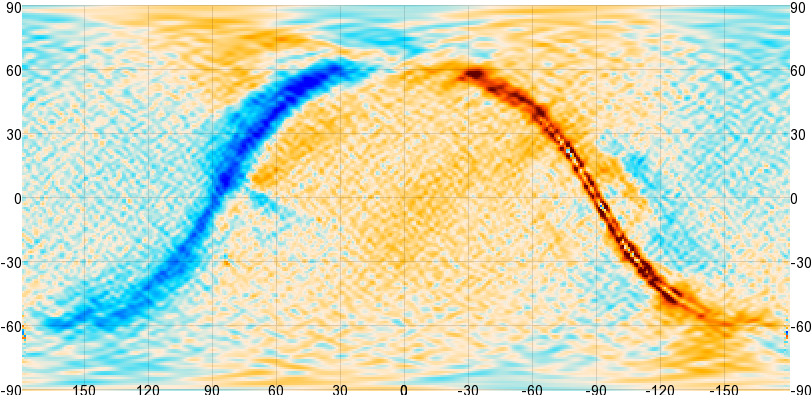} \\
		All &
		\includegraphics[height=28mm,clip,trim=7.5mm 8mm 7.5mm 0]{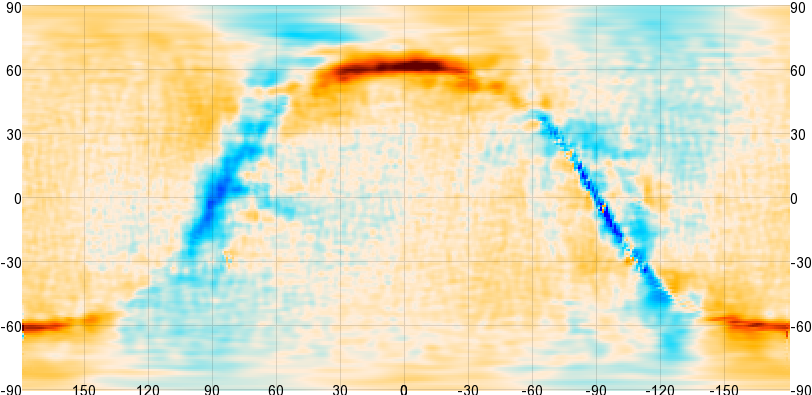} &
		\includegraphics[height=28mm,clip,trim=7.5mm 8mm 0 0]{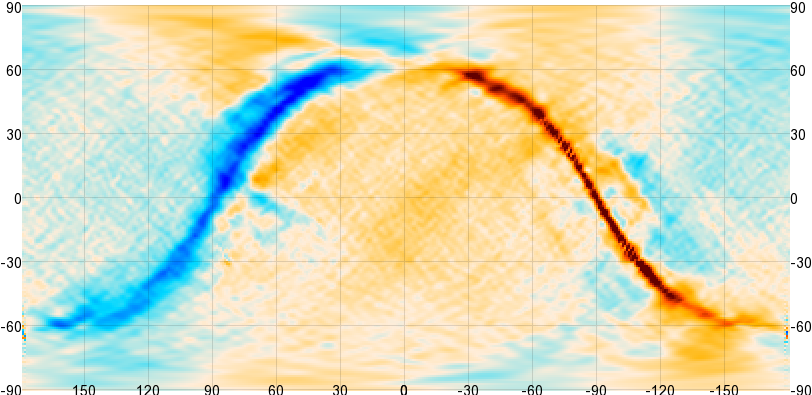}
	\end{tabular}
	\caption{As figure~\ref{fig:src_leak}, but for a cmb+dust simulation.
	\emph{Top}: Q and U maps based on only the Lx detector.
	The maps are dominated by O(1) leakage of small-scale I. \emph{Middle}:
	When coadding the Lx and Ly detectors the leakage is greatly reduced. If
	the two detectors have the same beam (as would be expected in the real instrument),
	then this cancellation would be perfect. In our simulation there is a 10\% mismatch,
	so the leakage is only suppressed by 90\%. \emph{Bottom}: Map based on all
	four detectors (Lx+Ly+Rx+Ry). Because we assumed the same beams for the detectors
	in the left and right horns, the leakage cancels. Leakage would only survive if
	both the detectors in a horn are mismatched \emph{and} the horns themselves are
	mismatched. The color scale is $\pm 400$ kJy/sr in I and $\pm 500$ Jy/sr
	in Q and U.}
	\label{fig:cmb_leak}
\end{figure}
To first order, PIXIE is immune to intensity to polarization
(I-to-P) leakage because any term sourced by I is not
modulated as a spin-2 field as the telescope spins, and is
therefore not classified as polarization. We confirm this
in simulations, where even 100\% I-to-P leakage in the
optics have no effect on the result.

However, this spin-separation of polarization can itself
become a source of polarization leakage.
If the telescope barrel is not perfectly aligned with PIXIE's
spin axis, then the beam will trace small circles in the
sky during each spin. Any local intensity quadrupole around
the point PIXIE is pointing at will show up as a spin-2
modulated signal in the time-ordered data, and will hence
be interpreted as polarization. Alternatively, the same
thing can happen if the barrel is correctly aligned,
but the beam is elliptical.

We investigated ellipticity-induced leakage for a
point source in figure~\ref{fig:src_leak} and for
a cmb + dust map in figure~\ref{fig:cmb_leak}, both
for a highly elliptical beam with a flattening of 0.5.
For each detector in isolation, the ellipticity results in a strong
quadrupolar leakage pattern that completely dwarfs the
intrinsic polarization. However, when combining
detectors to make a full map this leakage is doubly canceled.
Firstly, because the two detectors in a horn have the same
I response but opposite polarization response, the leakage
from each detector cancels. And secondly, the left and right
horn also differ by an overall sign in their polarization sensitivity,
leading to a second cancellation. All in all, I-to-P leakage
is a third order effect in PIXIE, and is unlikely to be an
important systemaic effect.

\subsection{Beam circularization}
\begin{figure}
	\centering
	\fboxsep=0mm
	\begin{tabular}{cccc}
		Input beam & Output I & Output Q & Output U \\
		\fcolorbox{border}{white}{\includegraphics[height=30mm]{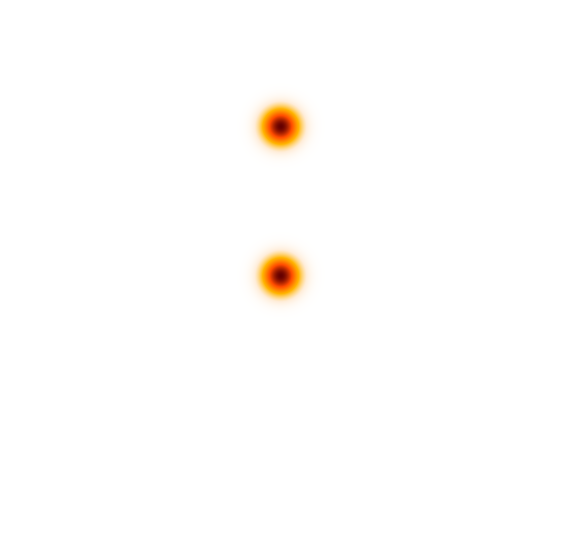}} &
		\fcolorbox{border}{white}{\includegraphics[height=30mm]{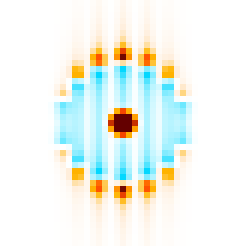}} &
		\fcolorbox{border}{white}{\includegraphics[height=30mm]{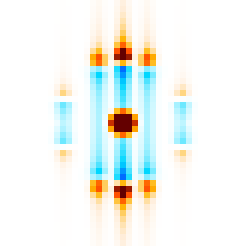}} &
		\fcolorbox{border}{white}{\includegraphics[height=30mm]{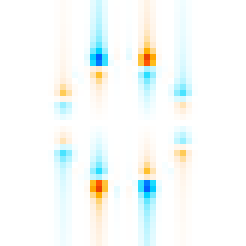}}
	\end{tabular}
	\caption{Beam circularization failure for a multimodal input beam. Each
	component has a FWHM of $1.9\degree$, and they are separated by $10\degree$.
	One would expect the off-center beam component to be smoothed into a smooth
	circle, but due to the limited amount of spin angles after orthogonalization
	the result is a dashed circle. Additionally, the sharp angular dependence
	of the beam results in interpolation failure during orthogonalization, which
	is responsible for the striping in the scanning direction. This failure mode
	is not relevant for realistic PIXIE beams.}
	\label{fig:badbeam}
\end{figure}
PIXIE's spin around the boresight also circularizes the beam. In the ideal
orthogonal case, where the telescope does not scan or stroke while spinning, every
pixel would be observed equally at every spin angle, leading to perfect
circularization. In practice, though, we need to rely on interpolation to
simulate an orthogonal scanning pattern, and this relies on the signal changing
smoothly as the spin angle changes in $22.5\degree$ steps (since there are 16
half-strokes per full spin)\footnote{This is for spin-2 orthogonalization.
With spin-1, we get 8 full strokes per spin and hence $45\degree$ steps.}
\footnote{While the autocorrelation function repeats
every quarter-stroke, every other quarter-stroke is mirrored, so they do not
form an equi-spaced dataset suitable for Fourier interpolation when orthogonalizing.
Otherwise, we could have gotten both 32 orthogonalized spin angles would have been
possible.}.
But a very long and thin beam could in theory sweep
over a small feature in far shorter time than this. This would create
complicated spatially dependent artifacts in the autocorrelation function.

Furthermore, even if the orthogonalization went perfectly, it would still only
result in 16 discrete, evenly spaced spin angles\footnote{When using
high-resolution mapping (see panel F in figure~\ref{fig:ortho}) this reduces to
8.} per pixel. The resulting beam is then the average over these 16
orientations, and so would not be perfectly radially symmetric, but would
instead have an 16-fold angular symmetry.

These effects are illustrated in figure~\ref{fig:badbeam}, for a
multimodal beam with two circular components separated by $10\degree$.
The resulting beam is not circular, and has quite complicated sub-structure.

How important are these limitations in practice? As we can see in
figure~\ref{fig:src_leak}, even for a beam with an unrealistically large
flattening of 0.5 any residual non-circularity is low enough
that it is drowned out by sub-pixel effects.

\subsection{Correlated noise}
\label{sect:corrnoise}
\begin{figure}
	\centering
	\hspace*{-13mm}\begin{tabular}{m{80mm}m{80mm}}
		\vspace*{2.0mm}\includegraphics[width=87mm]{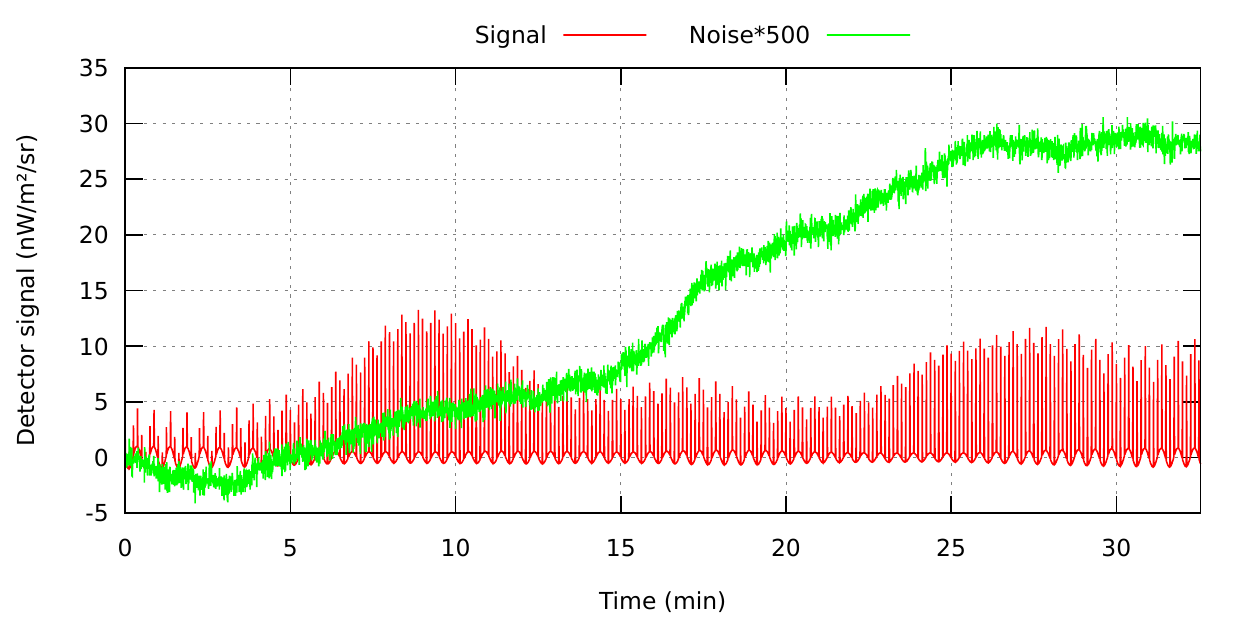} &
			\includegraphics[width=94mm]{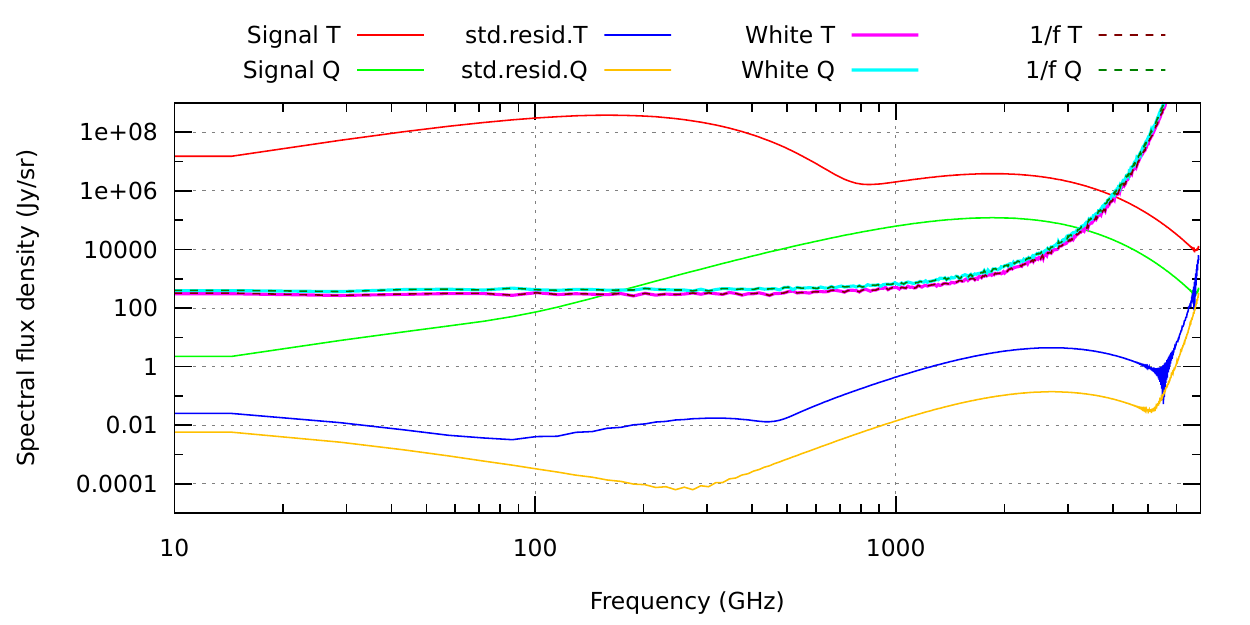}
	\end{tabular}
	\hspace*{-5mm}\begin{tabular}{m{56mm}m{54.4mm}m{56mm}}
		\includegraphics[height=28mm,clip,trim=0 8mm 7.5mm 0]{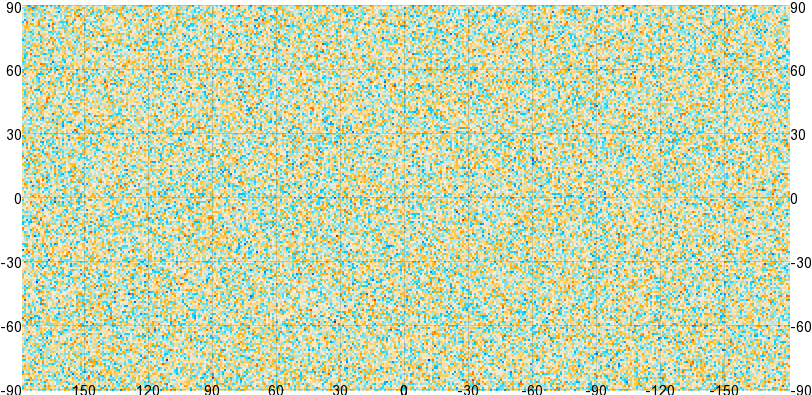} &
		\includegraphics[height=28mm,clip,trim=7.5mm 8mm 7.5mm 0]{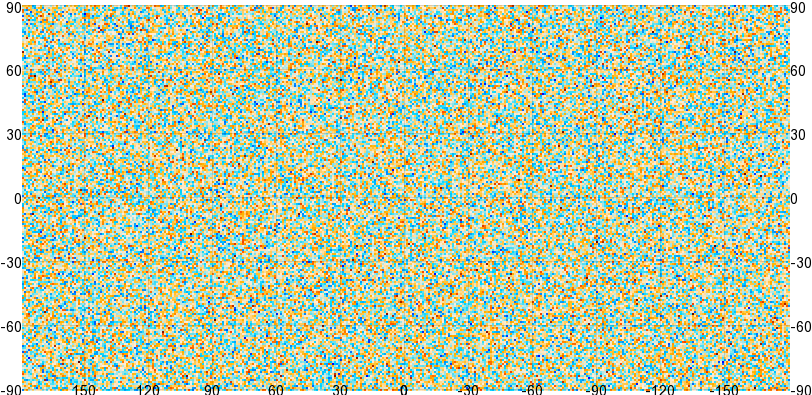} &
		\includegraphics[height=28mm,clip,trim=7.5mm 8mm 0 0]{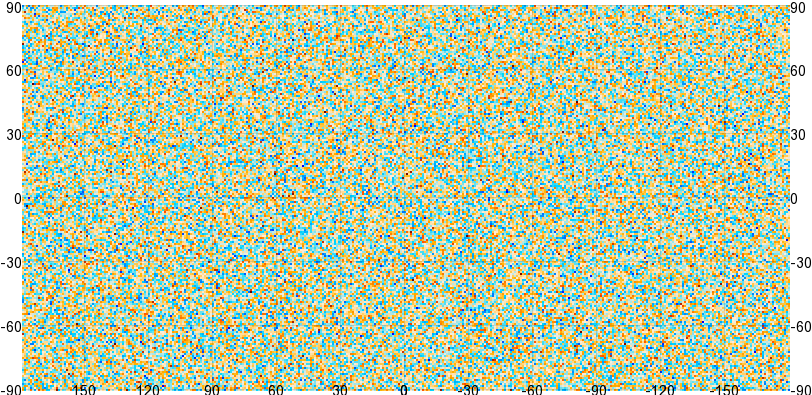} \\
		\hspace{32mm}I & \hspace{32mm}Q & \hspace{32mm}U
	\end{tabular}
	\caption{\emph{Top left}: Example noise realization with 1/f-noise with
		$f_\textrm{knee} = 0.1$ Hz, $\alpha=-3$ and $\sigma = 83 \sqrt{\rm s}$
		fW/$\textrm{m}^2$/sr per detector. This $f_\textrm{knee}$ is
		several orders of magntiude larger PIXIE's expected value,
		resulting in strongly non-white noise on $>10$ s time scales.
		\emph{Top right}: The resulting noise RMS is $92 \textrm{Jy/sr}$ per 15 GHz
		bin per ${1\degree}^2$ pixel in I and $128 \textrm{Jy/sr}$ in Q and U. These
		numbers take into account that our actual bins are slightly smaller than
		15 GHz, and that our pixels are slightly smaller than ${1\degree}^2$.
		Despite the high $f_\textrm{knee}$, the resulting spectrum noise is
		indistinguishable from white. The spectrum (and map) has been
		rescaled to correspond to 15 months of data. \emph{Bottom}: Map
		noise resulting from single barrel mapmaking with this noise sim.
		The color scale is $\pm 500$ \textrm{Jy/sr}. Like in the spectrum, the map noise
		is practically white.}
		\label{fig:corrnoise}
\end{figure}
PIXIE's detector readout is expected to have a slowly varying 1/f noise component,
with an $f_\textrm{knee} \sim 1/\textrm{hour}$ or longer. 1/f noise usually manifests as
correlated structures in the map, for example striping in the scanning direction.
In PIXIE's case, though, we get a measurement of the spectrum every 7.5 seconds
as the mirror strokes. As long as it is purely additive, slow drifts can be thought
of as low-order polynomials vs mirror position for a single mirror stroke. As such,
they are Fourier-transformed to low optical frequencies and affect only the DC component,
or at worst the first few bins of the synthesized spectra, and do not propagate to
spatial striping in the maps. We therefore expect PIXIE's
maps/spectra to have only white noise.

To test this we simulated 1/f noise with
power $(1+[f/f_\textrm{knee}]^\alpha)\sigma^2$, with $f_\textrm{knee} = 0.1 \textrm{Hz}$
(hundreds of times higher than expected, but slower than the stroke frequency),
$\alpha=-3$ and $\sigma = 83\sqrt{\rm s} \textrm{fW}/\textrm{m}^2/\textrm{sr}$ per detector.
As figure~\ref{fig:corrnoise} shows, this results in noise that white both
spatially and spectrally. This confirms our expectation that (additive) correlated
noise should not be an issue for PIXIE.

\subsection{Sub-pixel effects}
Anisotropy on scales smaller than PIXIE's beam, including CMB anisotropy
as well as point sources, will affect the map making. Scanning
the beams across the sky causes point sources to enter or exit the beam
during the course of a single mirror stroke, creating signal variations
on time scales short compared to the mirror stroke. Signals on short
time scales (high spatial frequencies) are Fourier-transformed to
high frequencies in the synthesized spectra and primarily affect channels
above 6 THz containing little true sky signal. See appendix~\ref{sect:subpix}
a fuller discussion.

\subsection{Mirror jitter}
\begin{figure}
	\centering
	\hspace*{-5mm}\begin{tabular}{ccc}
		\includegraphics[height=57mm,clip,trim=0 0 5mm 0]{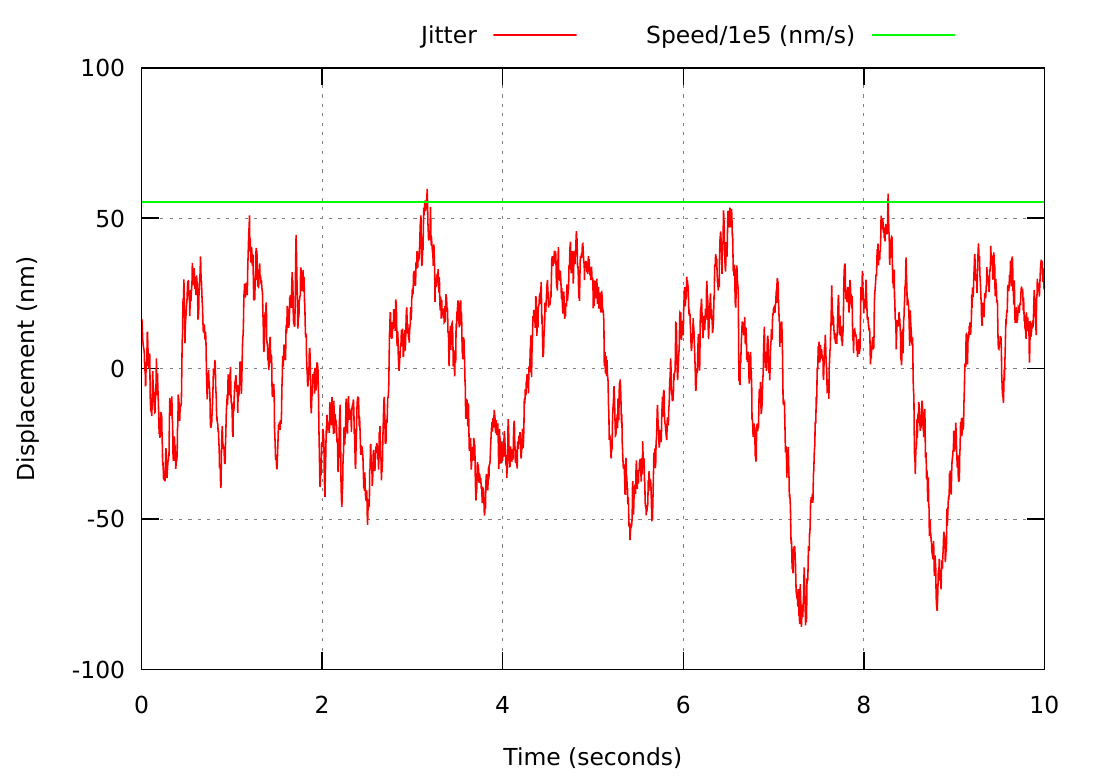} &
		\includegraphics[height=60mm,clip,trim=0 0 6mm 0]{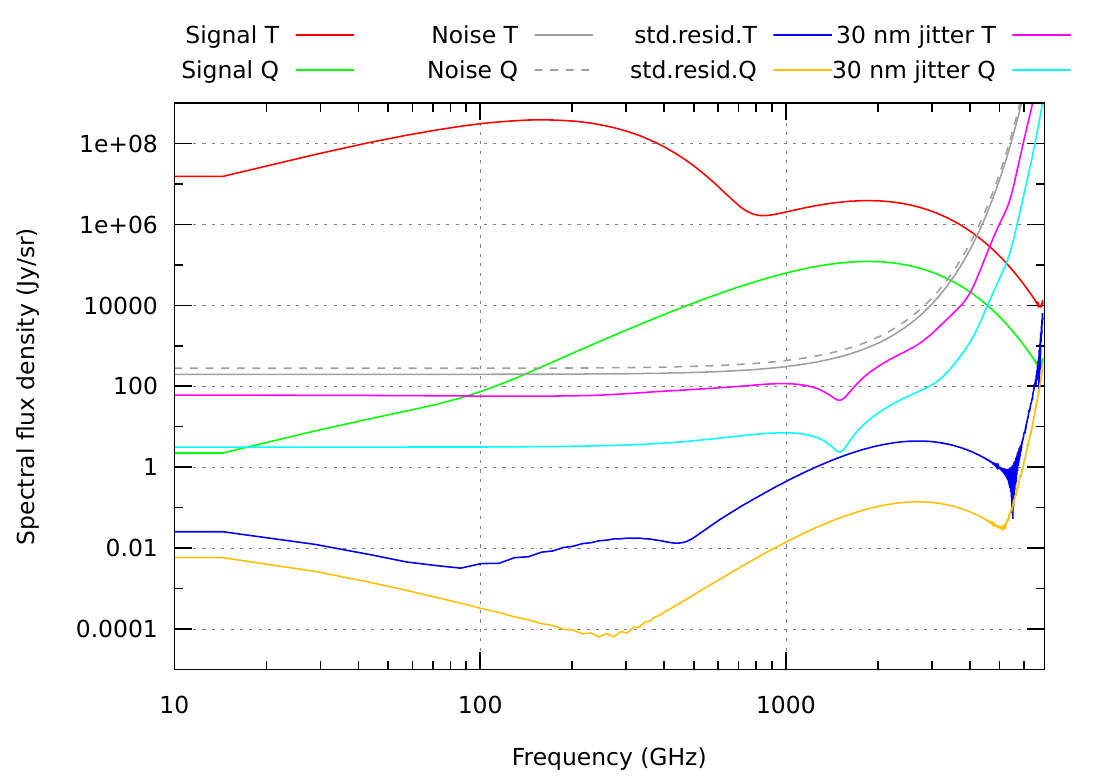}
	\end{tabular}
	\caption{The effect of jitter in the mirror position. \emph{Left}:
	We added noise consisting of a sum of 100 sine waves with random
	periods logarithmically distributed in the range 0.25 Hz to 1000 Hz
	with power proportional to $f^{-1}$ and a combined RMS of $30 \rm nm\sqrt{s}$.
	This is a rough approximation of noise that is damped
	on long timescales by the mirror control system and on short timescales
	by the inertia of the mirror. An example of such a noise realization
	is shown in red,
	and compared to the mirror average displacement per second divided by $10^5$.
	\emph{Right}:
	The average single-pixel spectrum error caused by this jitter
	compared to the signal, PIXIE noise and subpixel/subsample bias.
	A mirror jitter of the form and magnitude simulated here would be
	the largest of the systematic effects investigated in this
	paper. This plot is for single barrel mode, but the jitter noise
	curve in polarization is the same in double barrel mode.}
	\label{fig:jitter}
\end{figure}
In PIXIE's map-making we model the interferometer path delay as
changing at a constant rate in a triangle wave pattern. But in a
real experiment the mirror is a physical device that cannot turn
around instantly, and which will end up vibrating and jittering
at some level. The exact performance that can be expected from
the mirror is still uncertain, but according to \citet{pixie-systematics},
a jitter in the path difference of $\sim 30 \rm nm\sqrt{s}$ has been
achieved.

We simulated jitter at this level by adding noise consisting of a
sum of 100 sine waves with random periods logarithmically distributed
in the range 0.25 Hz to 1000 Hz with power proportional to $f^{-1}$
and a combined RMS of $30 \rm nm\sqrt{s}$ to the path delay in the simulator.
This is a rough approximation of noise that is damped on long timescales
by the mirror control system and on short timescales by the inertia of
the mirror. The resulting spectra are shown in figure~\ref{fig:jitter}.
Jitter both smoothes the autocorrelation function and acts as an extra
noise component, but as the figure shows the latter dominates for PIXIE.

The jitter shows up as an extra white noise component in the spectra,
and unlike the other effects we have investigated, this contribution to the
noise is not negligible. In total intensity it is about half as as high as
the expected instrument noise, while it's about two orders of magnitude
below the noise in polarization. The jitter level at which the jitter noise
and instrument noise are equal are $0.10 \rm \mu m\sqrt{s}$
for I,  $2.2\rm\mu m\sqrt{s}$ for P (single barrel) and
$3.1\rm\mu m\sqrt{s}$ for P (double barrel).

One might hope that double barrel mode would
be less affected by jitter noise, as it cancels the huge total intensity signal, and hence
could prevent it from contributing to this noise term. However,
the opposite polarization sign for the two detectors in each horn cancels
the total intensity contribution to the polarization jitter noise even in
single barrel mode. Mirror jitter is therefore approximately as important in double
barrel mode\footnote{Double barrel mode is more robust to I-to-P leakage,
so in the presence of uncancelled I-to-P leakage, the single barrel polarization
jitter noise would increase above the level indicated in figure~\ref{fig:jitter},
while the double barrel jitter noise would stay as it is. So in that case double
barrel mode would be an improvement.}.
Ensuring a low mirror jitter should be a high
priority in the PIXIE hardware design.

\section{Conclusion}
We have developed time-ordered data simulator and map-maker for the proposed
PIXIE experiment,
and used them to test the impact of subpixel bias, intensity to
polarization leakage, beam ellipticity/off-axis pointing, correlated
noise and mirror jitter. We find PIXIE to be remarkably robust against
all these effects, with the exception of mirror jitter, which is a
potential concern. At jitter levels above $0.10 \mu\textrm{m}\sqrt{\textrm{s}}$
for I or $3.1\mu\textrm{m}\sqrt{\textrm{s}}$ for P, the jitter rather
than detector performance becomes the limiting factor for the instrument's sensitivity.

This simulation framework was developed for PIXIE, but can be
adapted for any similar CMB satellite mission operating with a Fourier
transform spectrometer.

\section*{Acknowledgements}
%We would like to thank D. J. Fixsen for in-depth explanation
%about the inner workings of PIXIE; and T. Louis for helpful comments
We would like to thank T. Louis for helpful comments
during the writing of the article. The Flatiron Institute is supported by the
Simons Foundation.

\bibliographystyle{act_titles}
\bibliography{refs}

\pagebreak

\appendix

\section{Sub-pixel effects}
\label{sect:subpix}
As we saw in section~\ref{sect:results}, even ideal, noiseless
simulations result in low-level deviation from the input, as
seen in map-space in figure~\ref{fig:spatbias} and in the
spectrum in figure~\ref{fig:specbias}.

\begin{figure}[h!]
	\centering
	\hspace*{-13mm}\begin{tabular}{m{59mm}m{59mm}m{59mm}}
		\hspace{30mm}Input & \hspace{20mm}Interpolations & \hspace{30mm}Residual \\
		\includegraphics[height=43mm,clip,trim=0 0 0 0]{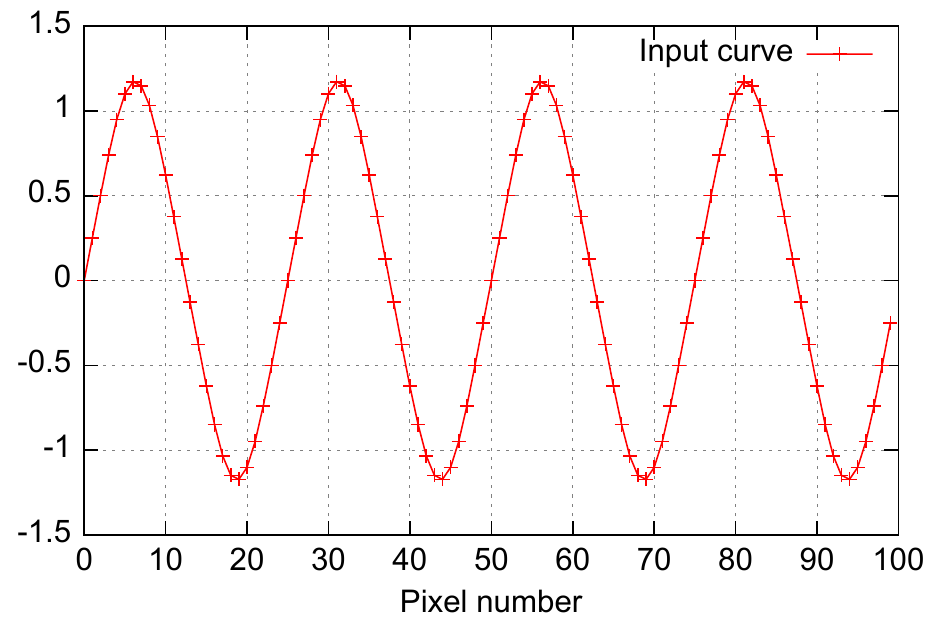} &
		\includegraphics[height=43mm,clip,trim=0 0 0 0]{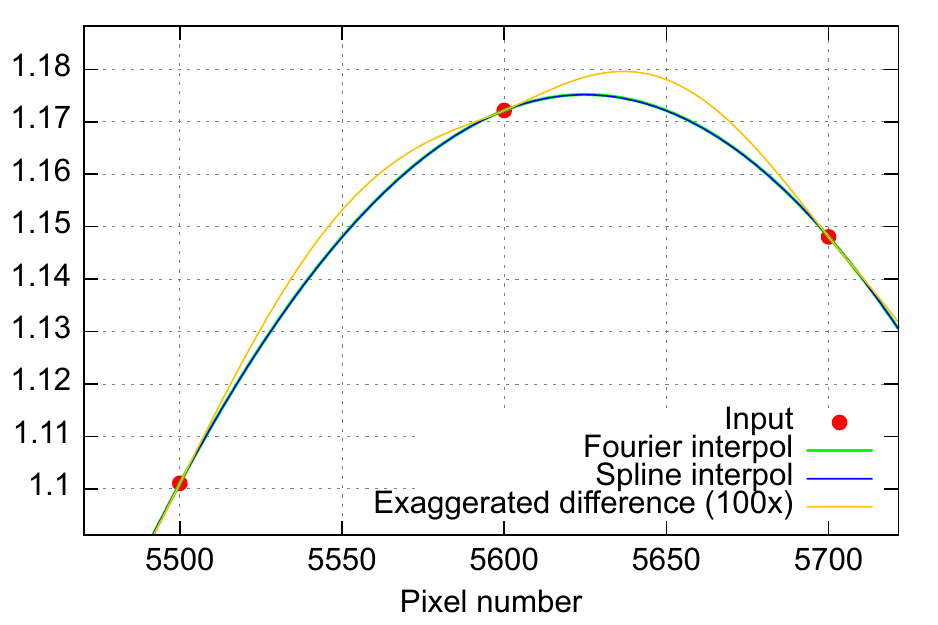} &
		\includegraphics[height=43mm,clip,trim=0 0 0 0]{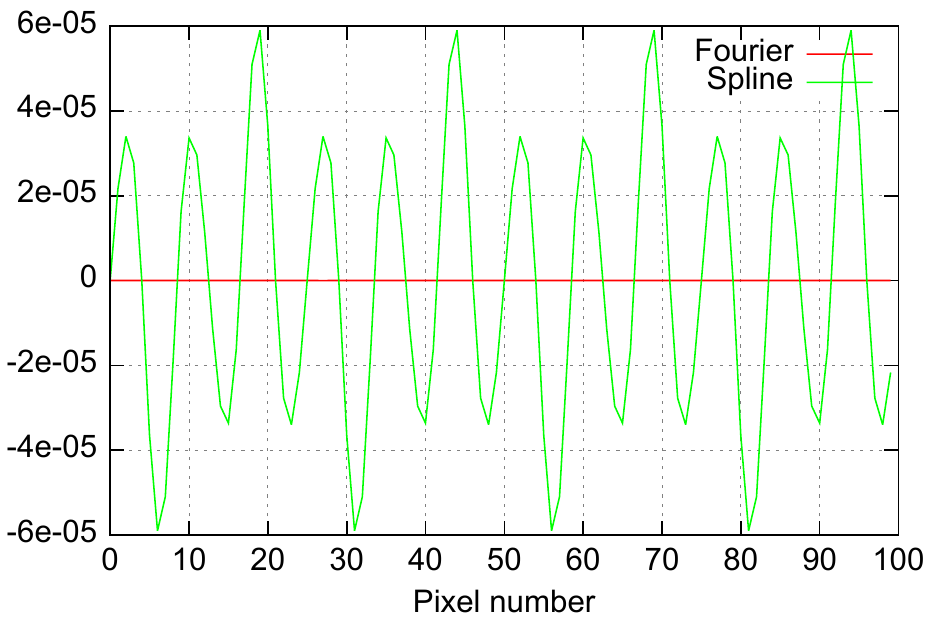}
	\end{tabular}
	\caption{1-dimensional toy example of sub-pixel mismatch bias.
	\emph{Left}: The input data set, which is a densely sampled sine wave with
	25 points per wavelength. \emph{Middle}: Two models for the sub-pixel behavior
	of the data set (Fourier and spline interpolation) disagree slightly.
	\emph{Right}: The residual after shifting the dataset right by half a sample
	(i.e. half-way between the red points in the middle panel) using Fourier
	(red) and spline (green) interpolation, and then back again using Fourier interpolation
	in both cases. When the forwards and backwards methods do not match, the result is biased.}
	\label{fig:iptoy}
\end{figure}

We believe that these biases are almost entirely due to sub-pixel effects.
During the orthoganalization procedure (section~\ref{sect:ortho})
interpolation is needed to go from the observed samples to the
idealized, orthogonal samples. The map-maker uses Fourier-interpolation
to do this, but the simulator uses bicubic spline interpolation to
generate samples that fall between input pixels. Hence, the
simulator and map-maker are making different assumptions about how
the signal behaves on sub-pixel scales, and this mismatch leads
to $\textrm{map}\rightarrow\textrm{tod}$ and $\textrm{tod}\rightarrow
\textrm{map}$ not being exact inverses of each other. Figure~\ref{fig:iptoy}
demonstrates this effect for a simpler 1-dimensional case.

Sub-pixel errors also occur during the sample window deconvolution.
The sample window is simulated by integrating sub-samples using
gaussian quadrature, which amounts to assuming that the sub-sample
behavior is described by a low-order polynomial. However, the
deconvolution is done by dividing by a sample window in Fourier space,
which assumes that the sub-pixel behavior is given by sines and
cosines bandlimited by the sampling frequency.

\begin{figure}
	\centering
	\hspace*{-2mm}\begin{tabular}{cc}
		I & Q \\
		\includegraphics[height=70mm,clip,trim=0 0 0 0]{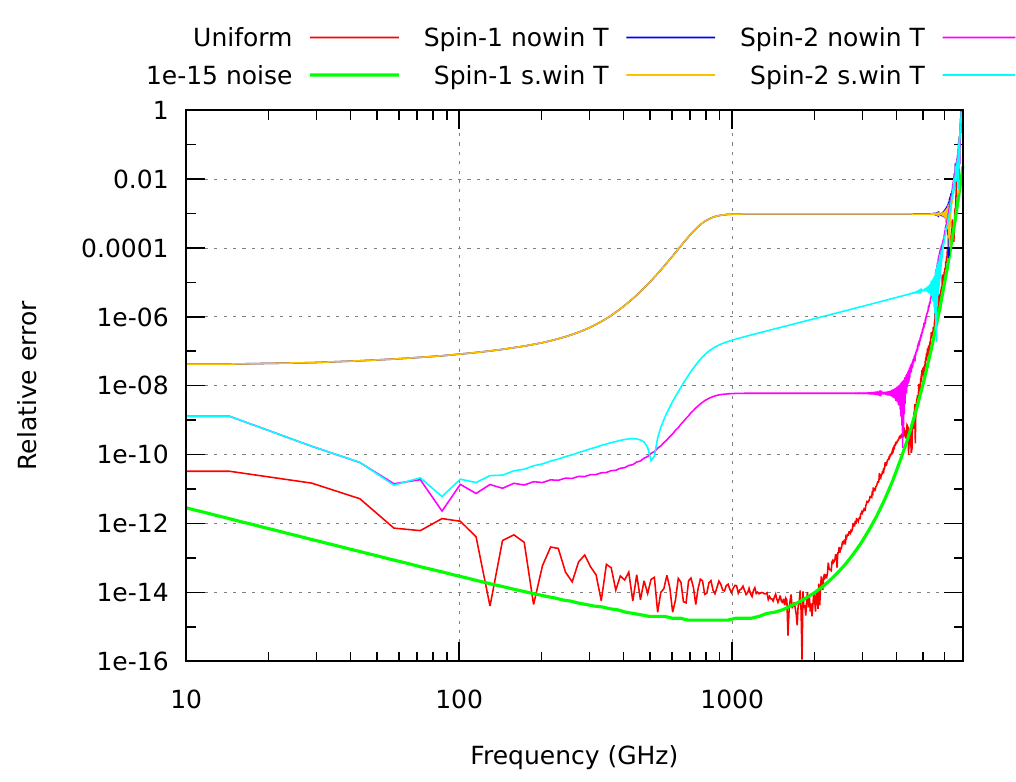} &
		\includegraphics[height=70mm,clip,trim=30mm 0 0 0]{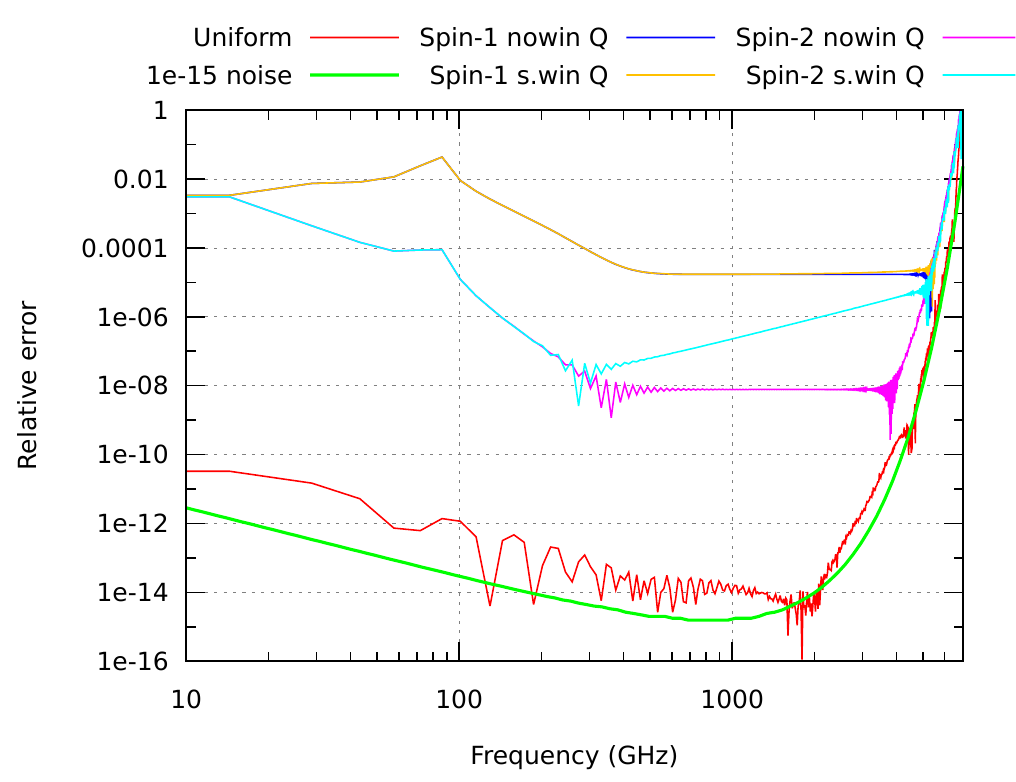}
	\end{tabular}
	\caption{The effect of sample window and Fourier shift distance on the
	mapmaking bias. Both of these are examples of sub-pixel bias. The sample
	window error comes from the mismatch between the gaussian quadrature that is
	used to integrate the sample window and the Fourier-space deconvolution that
	is used to remove it. It rises with frequency and is a $\sim 10^{-7}$ error
	at 1 THz. The Fourier shift error comes from the mismatch between the
	high-res bicubic sub-pixel behavior of the sky in the simulator and the
	bandlimited Fourier model used in the map-maker. It has a surprisingly large
	dependence on the interpolation distance. Spin-1 Fourier shifting has
	twice the interpolation distance of Spin-2 Fourier shifting, but $10^3-10^4$
	times as large a bias. The red curve shows the error for a
	monopole-only sky when using no sample window, demonstrating
	that aside from sub-pixel biases the accuracy is close to double
	precision float error.
	The relative Q error is high at low frequency because the signal itself
	becomes very small there. The feature at 90 GHz is due to Q changing
	sign there for this pixel.
	}
	\label{fig:subpixel-spec}
\end{figure}

We investigate the effect of these sub-pixel effects in
figure~\ref{fig:subpixel-spec}, where we first remove the
sample window (both in simulation and map-making), and then
make the map spatially homogeneous to eliminate sub-pixel effects.
After removing both of these, we are left with a relative error of
$\sim10^{-15}$, which we attribute to floating point errors.

These sub-pixel effects would not have appeared if the simulator
and map-maker both made the same assumptions about sub-pixel behavior,
which would have been the case for a simpler simulator.
That would, however, have been misleading. The real sky is neither
Fourier-interpolated nor bicubic-interpolated, and so this kind of
sub-pixel mismatch is unavoidable when analyzing actual data. The
mismatch between Fourier-interpolation and bicubic interpolation
is not exactly the same as what one can expect for the real data,
but it is a good approximation for it.

%\begin{figure}
%	\centering
%	\hspace*{-5mm}\begin{tabular}{cc}
%		I & Q \\
%		\includegraphics[height=80mm,clip,trim=0 0 6mm 0]{plots/spin_offset_T_abs_log_log.pdf} &
%		\includegraphics[height=80mm,clip,trim=27mm 0 0 0]{plots/spin_offset_Q_abs_log_log.pdf}
%	\end{tabular}
%	\caption{The interplay between having the barrel pointing offset
%	from the rotation axis and the spin of the Fourier shift in the
%	mapmaking algorithm for I (\emph{left}) and Q (\emph{right}).
%	Normally the spin-2 shift is much more accurate than the spin-1
%	shift, as can be seen by computing the green curve with its
%	dashed counterpart (see main text for discussion). This
%	is also the case when considering only the signal from a single
%	detector (blue). Adding a 1 arcsecond barrel offset slightly
%	increases the error consistent with a $\sim 1"$ smoothing
%	of the map (orange). However, if one only looks at a single
%	detector the error increase is dramatic (magenta) for
%	the spin-2 shift in Q. This is because the spin-2 shift assumes
%	that the signal is the same after a $180\degree$ spin rotation,
%	but with a barrel offset this is not the case. To a lesser extent
%	a barrel offset also turns the local quadrupole into $I\leftrightarrow P$
%	leakage, but this term has the opposite sign for oppositely aligned
%	detectors, and hence cancels.}
%\end{figure}

\end{document}